\documentclass[aps,onecolumn,preprint,prd,superscriptaddress,nofootinbib,floats]{revtex4}
\usepackage{amsmath,amssymb,color,mathrsfs,graphicx,verbatim,epsf,epsfig,bbm,url}
\usepackage[hyperfootnotes=false]{hyperref}
\usepackage{slashed}
\usepackage[config=altsf]{subfig}
\usepackage{multirow}
\allowdisplaybreaks

\def\beq{\begin{equation}}
\def\eq{\end{equation}}
\def\eeq{\end{equation}}

\def\centeron#1#2{{\setbox0=\hbox{#1}\setbox1=\hbox{#2}\ifdim
\wd1>\wd0\kern.5\wd1\kern-.5\wd0\fi
\copy0\kern-.5\wd0\kern-.5\wd1\copy1\ifdim\wd0>\wd1
\kern.5\wd0\kern-.5\wd1\fi}}
\def\ltap{\;\centeron{\raise.35ex\hbox{$<$}}{\lower.65ex\hbox{$\sim$}}\;}
\def\gtap{\;\centeron{\raise.35ex\hbox{$>$}}{\lower.65ex\hbox{$\sim$}}\;}

\def\chii0{\chi_i^0}
\def\chij0{\chi_j^0}

\def\foursqr#1#2{{\vcenter{\vbox{
 \hrule height.#2pt
 \hbox{\vrule width.#2pt height#1pt \kern#1pt
 \vrule width.#2pt}
 \hrule height.#2pt
 \hrule height.#2pt
 \hbox{\vrule width.#2pt height#1pt \kern#1pt
 \vrule width.#2pt}
 \hrule height.#2pt
     \hrule height.#2pt
 \hbox{\vrule width.#2pt height#1pt \kern#1pt
 \vrule width.#2pt}
 \hrule height.#2pt
     \hrule height.#2pt
 \hbox{\vrule width.#2pt height#1pt \kern#1pt
 \vrule width.#2pt}
 \hrule height.#2pt}}}}
\def\psqr#1#2{{\vcenter{\vbox{\hrule height.#2pt
 \hbox{\vrule width.#2pt height#1pt \kern#1pt
 \vrule width.#2pt}
 \hrule height.#2pt \hrule height.#2pt
 \hbox{\vrule width.#2pt height#1pt \kern#1pt
 \vrule width.#2pt}
 \hrule height.#2pt}}}}
\def\sqr#1#2{{\vcenter{\vbox{\hrule height.#2pt
 \hbox{\vrule width.#2pt height#1pt \kern#1pt
 \vrule width.#2pt}
 \hrule height.#2pt}}}}

\def\figin{\epsfcheck\figin}\def\figins{\epsfcheck\figins}
\def\epsfcheck{\ifx\epsfbox\UnDeFiNeD
\message{(NO epsf.tex, FIGURES WILL BE IGNORED)}
\gdef\figin##1{\vskip2in}\gdef\figins##1{\hskip.5in}
\else\message{(FIGURES WILL BE INCLUDED)}%
\gdef\figin##1{##1}\gdef\figins##1{##1}\fi}
\def\DefWarn#1{}
\def\figinsert{\goodbreak\midinsert}
\def\ifig#1#2#3{\DefWarn#1\xdef#1{fig.~\the\figno}
\writedef{#1\leftbracket fig.\noexpand~\the\figno}%
\figinsert\figin{\centerline{#3}}\medskip\centerline{\vbox{\baselineskip12pt
\advance\hsize by -1truein\noindent\footnotefont{\bf
Fig.~\the\figno:\ } \it#2}}
\bigskip\endinsert\global\advance\figno by1}

\def\fig#1#2#3#4{\vskip 0.5cm \begingroup \midinsert \centerline{
\psfig{file=#1,width=#2}} \vskip 0.4cm
\global\advance\figno by 1
\centerline{\vbox{\baselineskip=12pt \noindent Figure \the\figno: #3}}
\endinsert \endgroup {\xdef#4{\the\figno}} }

\def\figcrop#1#2#3#4#5#6#7#8{\vskip 0.5cm \begingroup \midinsert \centerline{
\psfig{file=#1,width=#2,bbllx=#3,bblly=#4,bburx=#5,bbury=#6}} \vskip 0.4cm
\global\advance\figno by 1
\centerline{\vbox{\baselineskip=12pt \noindent Figure \the\figno: #7}}
\endinsert \endgroup {\xdef#8{\the\figno}} \vskip .5cm}

\def\figlabel#1{\xdef#1{\the\figno}}
\def\encadremath#1{\vbox{\hrule\hbox{\vrule\kern8pt\vbox{\kern8pt
\hbox{$\displaystyle #1$}\kern8pt}
\kern8pt\vrule}\hrule}}
\def\underarrow#1{\vbox{\ialign{##\crcr$\hfil\displaystyle
 {#1}\hfil$\crcr\noalign{\kern1pt\nointerlineskip}$\longrightarrow$\crcr}}}

\begin{document}

\preprint{UTTG-23-13}
\preprint{TCC-019-13}
\preprint{FERMILAB-PUB-13-348-T}

\title{Improved mass measurement using the boundary of many-body phase space}

\author{Prateek Agrawal}
\email{prateek@fnal.gov}
\affiliation{Fermilab National Accelerator Laboratory\\
Batavia, IL, 60510}

\author{Can Kilic}
\email{kilic@physics.utexas.edu}
\author{Craig White}
\author{Jiang-Hao Yu}
\email{yujh@physics.utexas.edu}

\affiliation{ Theory Group, Department of Physics and Texas Cosmology Center \\
The University of Texas at Austin \\
Austin, TX 78712}

\begin{abstract}
We show that mass measurements for new particles appearing in decay
chains can be improved by determining the boundary of the available
phase space in its full dimensionality rather than by using
one-dimensional kinematic features for each stage of the cascade
decay. This is demonstrated for the case of one
particle decaying to three visible and one invisible particles in a
two-stage cascade, but our methods also apply to a more general set of
decay topologies. We show that not only mass differences, but also the
overall scale of masses can be determined with high precision without
having to rely on cross section information. The improvement arises
from the properties of the higher dimensional phase space itself,
independent of the matrix element for the decay, and it is not
weakened by the presence of intermediate on-shell particles in the
cascade. Our results are particularly significant for the case of low
signal statistics, a distinct possibility for new physics
searches in the near future.
\end{abstract}

\pacs{}
\maketitle

\section{Introduction}
The LHC experiments have collected and analyzed roughly 5~fb$^{-1}$ of
data for 7~TeV collisions and 20~fb$^{-1}$ of data for 8~TeV
collisions, with no obvious evidence of physics beyond the Standard Model (SM).
In many cases, limits on new colored states reach beyond a TeV.
It is then important to consider the possibility that
colored new states appear at higher energies than expectations based on minimal
tuning would imply. Mass limits on color-neutral final states are weaker,
but these are also produced with a lower cross section.
Therefore, even with the high energy run of the LHC it is quite possible that
if we encounter physics beyond the SM, we will only have a limited
number of signal events to study it. More quantitatively, with 100
fb$^{-1}$ of integrated luminosity at the LHC with 14 TeV collisions,
one would produce only $\mathcal{O}(10^3)$ events with pairs of 1 TeV stops
\cite{Beenakker:1997ut}, or $\mathcal{O}(10^4)$ events from electroweak pair production
of 300 GeV charginos/neutralinos \cite{Beenakker:1999xh}. Moreover,
the number of events will decrease further due to selection cuts and
branching fractions. Therefore, optimizing measurements such as mass
determination for low statistics scenarios becomes crucial. The goal
of this paper is to demonstrate that in cascade decays this can be
achieved by determining the boundary of the kinematically accessible
phase space in its full dimensionality.

The collider phenomenology of most extensions of the SM is dominated
by a  ${\mathbb Z}_2$ symmetry. Once ${\mathbb Z}_2$-odd particles are produced in pairs, they
will decay through a sequence of two and three-body decays until the
lightest ${\mathbb Z}_2$-odd particle is reached, which is stable and
escapes detection. Resonances are not directly observable in such
decay chains, so for the determination of the masses in the decay
chain one has to rely on kinematic edges and endpoints, maximal values
in the distribution of $(p_{i}+p_{j})^2$, where $p_{i}^{\mu}$ and
$p_{j}^{\mu}$ are the four-momenta of visible particles emitted in
subsequent stages of the decay. The position of each such kinematic
feature potentially offers a nontrivial relation between the masses of
the intermediate particles in the relevant stage of the decay chain.
Aside from matrix-element effects such as spin correlations, the
stages of the decay occur independently, and therefore these kinematic
features are studied independently for each stage in the cascade
decay.

An important weakness of kinematic edges and endpoints, the simplest one-dimensional kinematic observables, is that while they are sensitive to the differences between the unknown masses in the spectrum, they have a ``flat direction'' along the overall scale of masses. In other words, as all masses in the spectrum are raised and lowered together, edges and endpoints have reduced sensitivity. While using the cross section information may be used to gain additional sensitivity to the overall mass scale, this approach is often unreliable due to unknown theory parameters, especially when the underlying model for the new physics is not known in full detail. In this paper we will show that the method of using the full phase space density can yield a more precise determination of the overall scale of masses compared to kinematic edges and endpoints.

Many methods have been developed over the years for the
determination of masses of unknown particles in cascade decays.
Beyond the simplest edge/endpoint methods~\cite{Hinchliffe:1996iu,
Bachacou:1999zb,
Allanach:2000kt,
Lester:2001zx,
Gjelsten:2004ki,
Gjelsten:2005aw,
Lester:2005je,
Miller:2005zp,
Lester:2006yw,
Ross:2007rm,
Tovey:2008ui,
Barr:2008ba,
Barr:2008hv,
Webber:2009vm} these include polynomial methods
\cite{
Hinchliffe:1998ys,
Nojiri:2003tu,
Kawagoe:2004rz,
Gjelsten:2006as,
Cheng:2007xv,
Nojiri:2008ir,
Cheng:2008mg,
Cheng:2009fw,
Matchev:2009iw,
Autermann:2009js,
Kim:2009si,
Han:2009ss,
Kang:2009sk,
Webber:2009vm,
Nojiri:2010dk,
Kang:2010hb,
Hubisz:2010ur,
Cheng:2010yy,
Gripaios:2011jm,
Han:2012nm,
Han:2012nr}, and the use of
observables that are not fully Lorentz-invariant but are nevertheless
invariant under boosts along the beam direction
\cite{
Lester:1999tx,
Barr:2003rg,
Meade:2006dw,
Baumgart:2006pa,
Matsumoto:2006ws,
Lester:2007fq,
Cho:2007qv,
Gripaios:2007is,
Nojiri:2008hy,
Tovey:2008ui,
Cho:2008cu,
Serna:2008zk,
Barr:2008ba,
Nojiri:2008vq,
Cho:2008tj,
Cheng:2008hk,
Choi:2009hn,
Matchev:2009ad,
Cohen:2010wv,
Polesello:2009rn,
Konar:2009wn,
Konar:2009qr,
Curtin:2011ng,
Nojiri:2010mk,
Lester:2011nj,
Mahbubani:2012kx}, as these
are very useful in a hadron collider environment where the
longitudinal motion of the center of mass frame cannot be determined
(though this may not be essential for mass measurement
\cite{Agashe:2012bn, Agashe:2013eba}).
In mass measurements for known particles such as the top quark, or for
specific new physics theories, the
matrix element for the decay can yield additional information to
increase the precision of the measurement \cite{Kondo:1988yd,
Dalitz:1991wa,
Abazov:2004cs,
Artoisenet:2010cn,
Alwall:2010cq,
Fiedler:2010sg,
Gainer:2013iya,
Alwall:2009sv}. However, this is often computationally prohibitive.
There are many additional refinements of these ideas, some of which also
contain additional information about the overall mass scale in the
cascade (for instances in the context of the $M_{T2}$ variable, see \cite{
Cho:2007dh,
Barr:2007hy,
Barr:2009jv,
Cho:2007qv,
Barr:2007hy,
Burns:2008va,
Burns:2009zi,
Cho:2009zza} and references therein).

The reader will note that some of the references cited above do indeed
have sensitivity to the overall mass scale.  It is certainly possible
to design one-dimensional observables that are {\it optimized} for
measuring the overall mass scale. Often, this utilizes the same type
of correlations in the full phase space distribution as the
multi-dimensional analysis we propose. In this paper, we
will not study any optimized observables but instead we will juxtapose
the simplest one-dimensional kinematic observables (kinematic
edges and endpoints) with an equivalently simple multidimensional
analysis.  It will be interesting to compare the full phase space
analysis with existing optimized analyses, and to explore the extent
to which a multidimensional phase space analysis can be further
optimized.  We leave this for future work.

The properties of two- and three-body phase space relevant for decays
are very well understood and are discussed in textbooks.  The
Lorentz-invariant two-body phase space is trivial. The three body
phase space can be expressed, up to rotational invariance\footnote{It
is assumed here that the spin of the initial particle is averaged
over, as we wish to consider properties of the phase space itself.
However we will come back to the role of the decay matrix element
later in this section.}, in terms of two Lorentz invariants, which can
be taken to be $m_{12}^{2}$ and $m_{13}^{2}$ where
$m_{ij}^{2}=(p_{i}+p_{j})^2$. As is well known, this parametrization
has the additional useful property that the phase space density is
constant, \begin{equation} d\Pi_{3}={\rm const.}\times dm^{2}_{12}\
  dm^{2}_{13}.  \label{eq:dLIPS3} \end{equation}

In this paper we will show that when cascade decays are studied
stage-by-stage in this fashion, taking advantage of one-dimensional
kinematic features at each stage, a certain amount of information is
still left unused, namely phase space correlations between different
stages in the cascade, which can be used to improve the precision of
mass measurements especially at low statistics. We should emphasize
here that the correlations in question have to do with the
distribution of events in the full-dimensional phase space itself, not
with the matrix element of the decay. We will demonstrate further that
using the full dimensionality of phase space for mass determination
can yield information on the overall scale of masses rather than only
the difference of masses in the spectrum, a significant improvement
over methods relying on edges and endpoints.


Unlike the three-particle phase space, which has a constant volume
element when expressed in terms of the conventional Lorentz-invariant
parametrization, the corresponding volume element for $n$-particle phase space will
in general be a function of the phase space coordinates. The crucial
point that will be made quantitatively in the next section is that
when $n=4$, the volume element is actually enhanced near the boundary
of the kinematically allowed region\footnote{While this statement
still holds in the case of $n>4$, the mathematical details are more
subtle and will be studied in future work.}. This is ideally suited
for a mass measurement, which is mainly an attempt
at determining the boundary of kinematically accessible phase space,
so a phase space density
that is enhanced near the boundary means that a modest number of
events can still give a good resolution for the boundary. The enhancement in
question is not manifest in the more
familiar cartesian parametrization of phase space, namely
\begin{equation}
  d\Pi_{n}=\left(\prod_{i}
  \frac{d^{3}p_{i}}{(2\pi)^{3}2E_{i}}\right)\delta^{4}(p_{0}^{\mu}-\sum_{i}p_{i}^{\mu}).
\end{equation}
Furthermore, even when the cascade decay contains intermediate
particles that go on-shell, we will show that the enhancement still
persists, but only when the full cascade is analyzed at once, thus
keeping correlations between different stages of the cascade.
Analyzing each stage of the cascade independently by looking for
kinematic features in several one-dimensional distributions such as
edges and endpoints (henceforth referred to simply as an ``endpoint
analysis''), does not utilize the full information available in these
correlations.

\begin{figure}[t]
  \centering
  \includegraphics{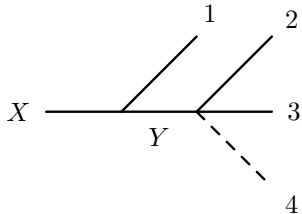}
  \caption{A decay topology where the mother particle $X$ undergoes a
  two body-decay to the intermediate on-shell particle $Y$, which
  subsequently decays through an off-shell state to the stable
  invisible particle $4$. The particles associated with momenta
  $p_{1,2,3}$ are assumed to be observable SM particles.}
  \label{fig:dalitz1}
\end{figure}

\begin{figure}[t] \centering
  \includegraphics[width=2.0in]{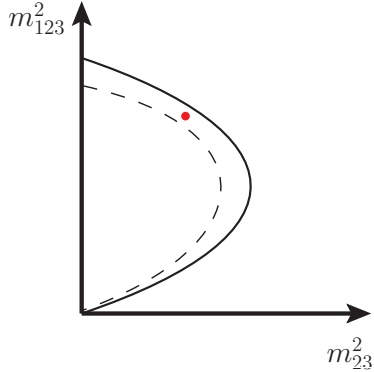}
  \caption{A two-dimensional slice of phase space for the decay
  topology of figure~\ref{fig:dalitz1}. The solid curve indicates the
  boundary of the kinematically accessible phase space, and the dashed
  curve indicates the boundary obtained using an incorrect hypothesis
  about the unknown masses. The single dot stands for a single decay
  event.} \label{fig:dalitz2} \end{figure}

Let us demonstrate this point in a specific example, namely the decay
topology seen in figure~\ref{fig:dalitz1}. In terms of one-dimensional
invariants, one expects an endpoint in $m_{23}$, as well as an edge in
$m_{123}$ for each fixed value of $m_{23}$\footnote{For a fixed value
of $m_{23}$, the $23$-system can be regarded as a single particle with
a fixed mass, which, combined with particle 1 produces a kinematic
edge because the intermediate particle $Y$ is assumed to be
on-shell.}. Now consider the single event depicted in
figure~\ref{fig:dalitz2}. The solid curve in this figure indicates the
true boundary of the kinematically allowed phase space while the
dashed curve indicates the boundary obtained using an incorrect
hypothesis for $m_{X}$, $m_{Y}$ and $m_{4}$. Two facts become
evident by considering this figure, namely that:
\begin{enumerate}
  \item The single event by itself is consistent with both mass
    hypotheses when the $m_{123}^2$ and the $m_{23}^2$ distributions
    are considered separately. However, it rules out the incorrect
    hypothesis when the full dimensionality of the phase space
    boundary is considered. Thus there is information in the full
    phase space distribution that is not available to either one of
    the one-dimensional projections.
  \item The event in question does
    not lie close to the endpoints in the projected distributions onto
    either the $m_{23}^2$ or the $m_{123}^2$ directions. Therefore
    this event will carry little weight in an endpoint analysis.
    In the case of a modest
    number of total events, the endpoints of any one-dimensional
    projection will be poorly populated while the full four-body phase
    space boundary will be well populated, leading to a more precise
    mass determination.
\end{enumerate}

The rest of this paper will be devoted to quantitatively studying this
improvement in the precision of mass measurements at low statistics.
We will do this by setting up a comparison between mass measurements
for the same underlying physics using multiple endpoint analyses and
a single multidimensional analysis (henceforth denoted as ``phase
space'' analysis). In order to
focus on the core issue, we will set up this comparison such that
many complications that would arise in a realistic study are removed.
We will argue below that these simplifying assumptions are not
expected to give an advantage to either method of mass measurement and
therefore a significant difference in the performance of the methods
should persist even as one moves toward a more realistic analysis.
The assumptions we make as we set up our analysis are as follows:

\begin{figure}[htp] \centering \subfloat[]{
  \includegraphics[width=0.4\textwidth]{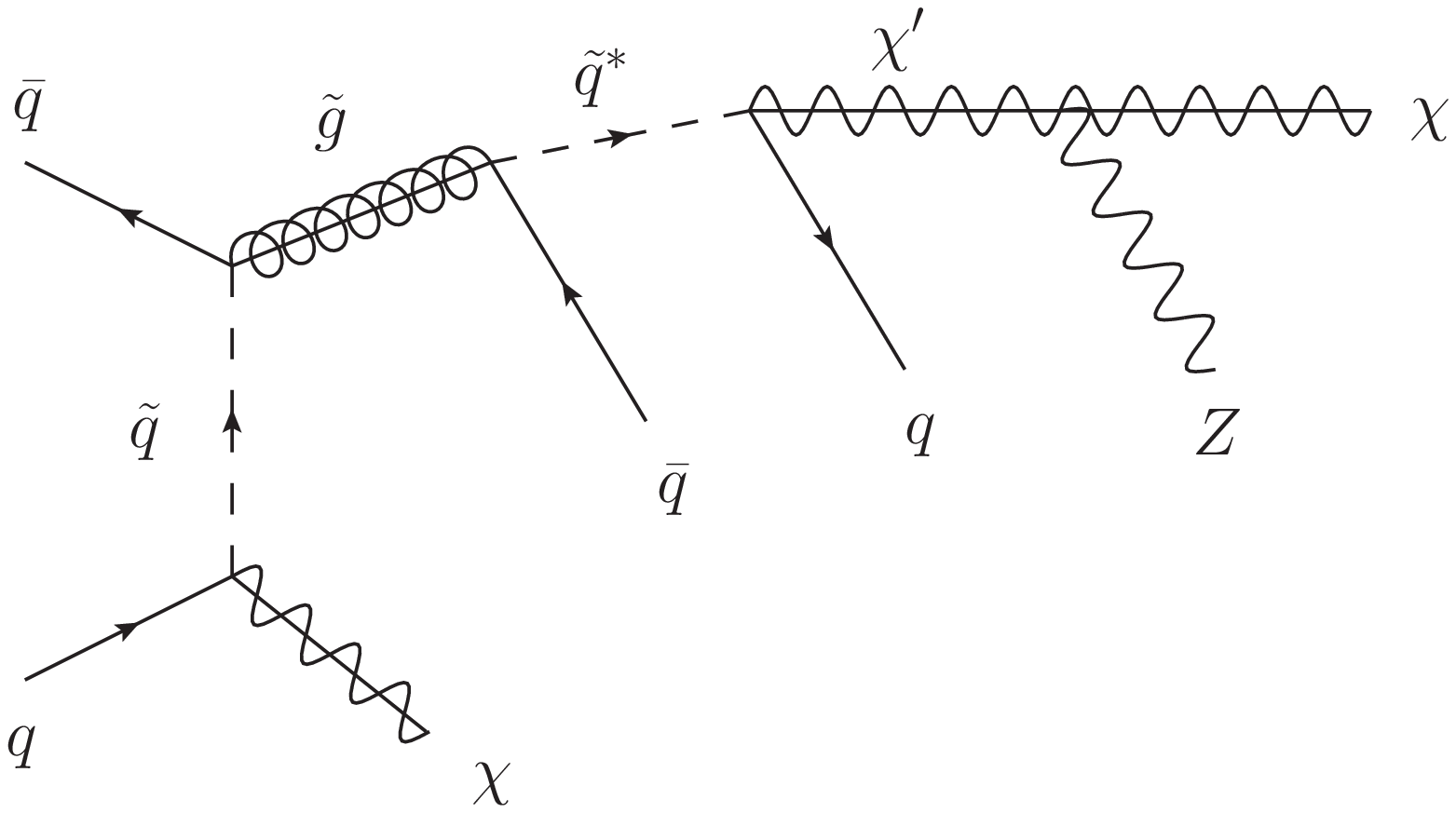}
  \label{fig:topo_gluino} } \qquad \subfloat[]{
  \includegraphics[width=0.4\textwidth]{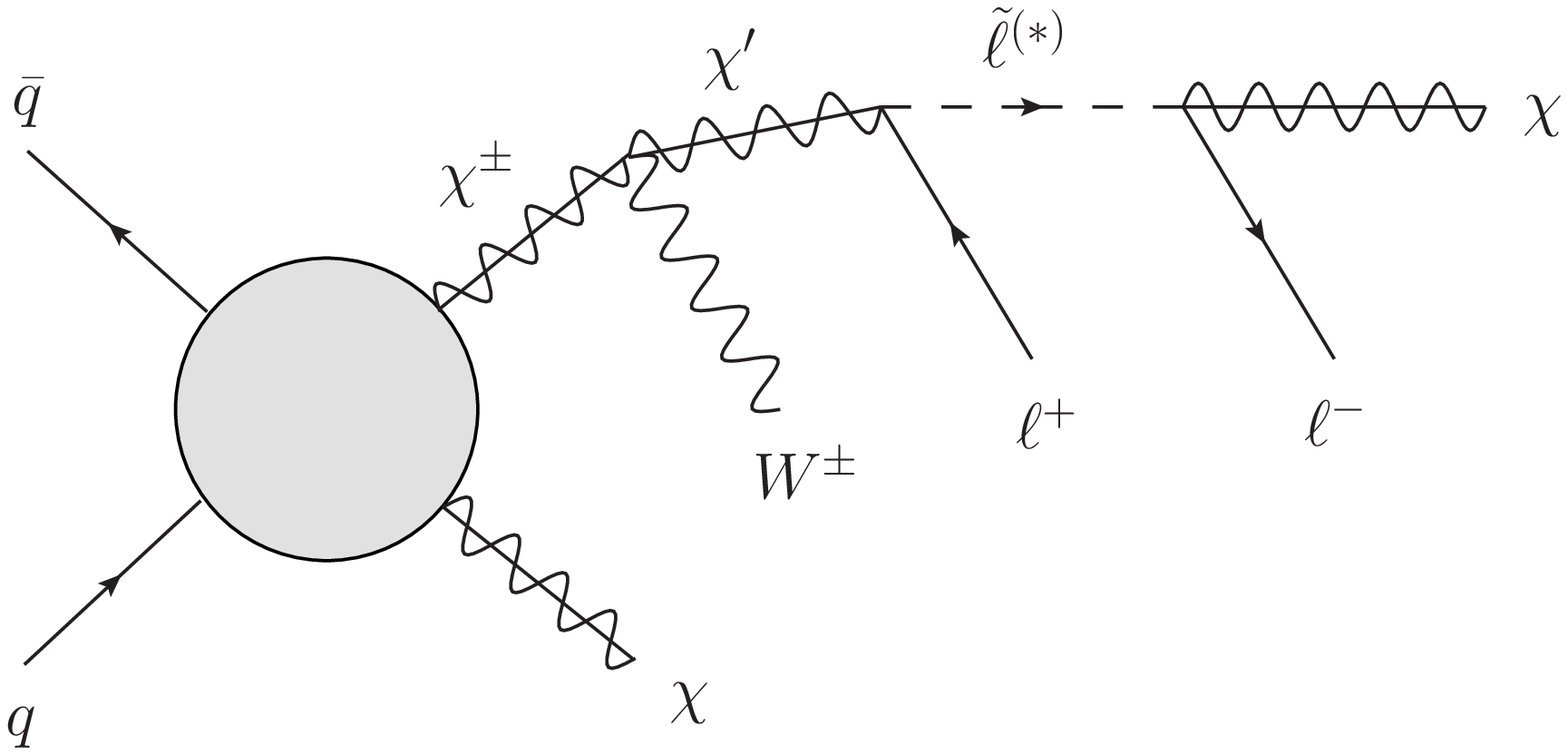}
  \label{fig:topo_chargino} } \\ \subfloat[]{
  \includegraphics[width=0.4\textwidth]{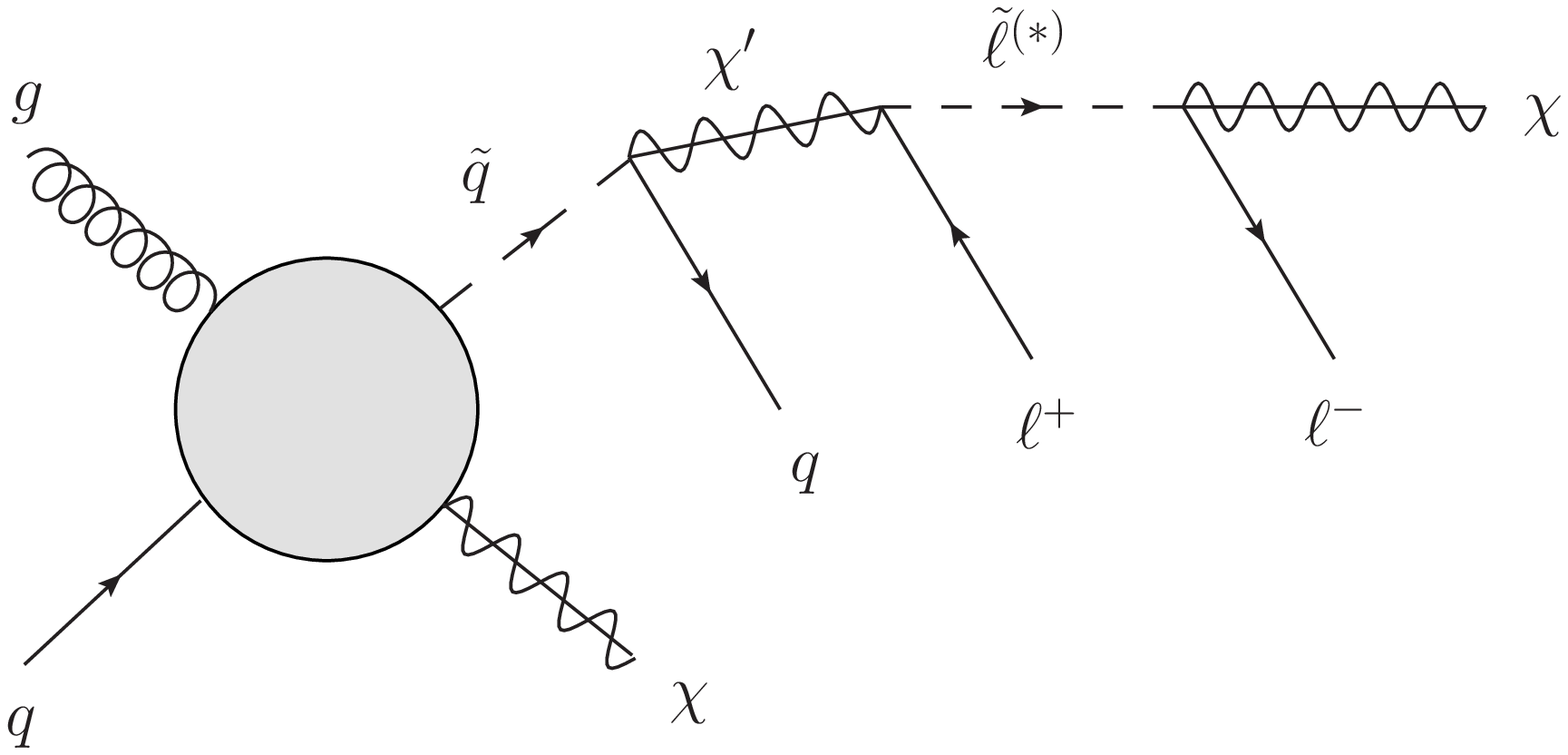}
  \label{fig:topo_squark} } \caption{Production mechanisms in
  supersymmetric scenarios for which the formalism presented in this
  paper will be directly applicable.} \label{fig:topo_all}
\end{figure}

\begin{itemize}

\item For the majority of beyond the SM (BSM) scenarios,
  the ${\mathbb Z}_2$ ensures that the particles in the new sector are
  produced in pairs rather than singly. In order to present as simple
  an implementation of our methods as possible, we will focus on models
where one of the produced particles is the lowest-lying stable
particle (LSP) while the other side involves a longer cascade. This choice
also reduces the combinatoric background. There are a variety of
motivated BSM processes for which such final states are relevant. Here
we mention a few examples from supersymmetry:

\begin{enumerate}
  \item
    A spectrum where all scalars super-partners are heavy, and the
    gluino is light enough to be singly produced in association with
    the LSP, but too heavy for the pair production to be observable.
    This is illustrated in figure \ref{fig:topo_gluino}. Such a
    spectrum can for instance arise in models of split supersymmetry
    \cite{
    ArkaniHamed:2004fb,
    Giudice:2004tc,
    ArkaniHamed:2004yi,
    ArkaniHamed:2012gw}.
  \item A spectrum where all colored states are
    heavy, but where chargino-neutralino production can allow one to
    access the electroweak sector. This is illustrated in figure
    \ref{fig:topo_chargino}. Here we assume that the chargino is
    moderately heavy such that the $W$ produced in the cascade is
    somewhat boosted and can be reconstructed as a single particle.
    This spectrum can also arise in moderately split models of
    supersymmetry.
  \item A spectrum where the gluino is very heavy,
    but squarks are light enough to be produced singly in association
    with the LSP. If there are further electroweak states between the
    squark and the LSP, one can easily get longer cascades. This is
    illustrated in figure \ref{fig:topo_squark}.
\end{enumerate}

In contrast to these asymmetric event topologies, pair production of a
heavy particle with long cascade chains on both sides of the event may
allow for the events to be fully reconstructed \cite{Cheng:2010yy}, or
failing that, it is still possible to measure all unknown masses using
the methods of ref.~\cite{
Cheng:2007xv,
Cheng:2009fw,
Matchev:2009iw,
Autermann:2009js,
Webber:2009vm,
Nojiri:2010dk}. However, we concentrate on an event
topology in this paper that is not easily handled by either one of
these methods.

Finally, there are additional subtleties in the analysis we present
when there are cascades on both sides of the event, as additional
Lorentz-invariant observables become relevant that are constructed
from the momenta coming from opposite sides of the event, thus the
phase space distributions of the two mother particles become further
correlated with one another. While these subtleties can indeed be
addressed within our formalism, this will be left for future
work.

\item In order to highlight the importance of the phase space density
in the mass measurement, we will neglect the effects of spin and
assume that all particles in the decay chain decay isotropically.
Clearly, this will not be the case for motivated extensions of the SM.
We will mention in section \ref{sec:conclusions} however how spin
information could be incorporated for specific models, and may actually
be used for the determination of all spins in the cascade, similar to
the original Dalitz analysis. In this paper we focus on the
determination of the boundary of phase space, while spin determination
will deal with the distribution of events throughout phase space, not
just the events close to the boundary. The simplifying assumption we
make in this paper will allow us to separate these two issues and
address the effects of spin in future work.

\item We will assume an ideal detector with perfect energy
resolution. This is an unrealistic assumption for final state jets,
however if the final state particles are leptons or photons, the
approximation may be acceptable.  In section \ref{sec:conclusions} we
will comment on how the effects of finite energy resolution can be
incorporated into the analysis described in section \ref{sec:methods}.

\item We will assume that the final state particles are
distinguishable. Since we will be dealing with one side of the event
only, which contains three visible particles, this assumption is not
entirely unrealistic. Nevertheless, we will comment in section
\ref{sec:conclusions} how the analysis may be modified in order to
take combinatoric backgrounds into account.

\item We will assume that the analysis starts out with the correct
hypothesis for the decay topology. This is only a mild assumption, as
there are only a few possible topologies in any case. Since our
analysis relies on likelihoods, it can in fact be used to compare the
quality of fit of alternative hypotheses (see also \cite{Bai:2010hd,
Birkedal:2005cm,
Costanzo:2009mq,
Rajaraman:2010hy,
Baringer:2011nh,
Choi:2011ys}).
Indeed, multibody phase space correlations are ideal for uncovering
the presence of intermediate on-shell particles, since in the
narrow-width approximation only certain (hyper)planes of the full
phase space will be populated, which can be used to identify the
underlying decay topology.

\item We will work with a signal-only sample. Clearly, in any
realistic analysis, we will need to include SM processes as background that can
mimic the relevant final state. Nevertheless, the LHC has already
pushed spectra of BSM models to high enough masses that it is
plausible to expect that analysis cuts on the transverse momentum of
final state particles can yield signal samples with high purity.
Furthermore, the likelihood-based analysis we present will not be
significantly affected by the presence of such analysis cuts, whereas
traditional methods that rely on finding kinematic edges and endpoints
may be biased when such cuts are made.

In any case, it is relatively straightforward to modify our methods to
account for the possibility that any given event may have been
produced from a background process, and this will be addressed in
future work. There is no reason to expect the background to exhibit an
enhancement near the boundary of phase space for the signal. Thus the
background distribution will be smoothly varying at the boundary of
the kinematically allowed phase space, where the signal sample will
have an enhancement. This should minimize the impact of background on
the applicability of our methods.

\end{itemize}

We begin by describing in section \ref{sec:ps} the Lorentz-invariant
parametrization of phase space for more than three final state
particles that is relevant for the rest of our analysis. Then, in
section \ref{sec:methods} we describe in detail how we set up the mass
measurement, both for the phase space and the
endpoint analysis that we use as a reference. We present the
results of our analysis for the case of a specific benchmark spectrum
in section \ref{sec:results}. We conclude in section
\ref{sec:conclusions}, and discuss future directions.

\section{Description of Four-Particle Phase Space}
\label{sec:ps}
Our claim that higher precision in mass measurements is possible relies strongly on the
fact that the boundary of kinematically accessible phase space is well
populated for a single particle decaying to four final states. In order to show that this is indeed the case, this section will be devoted to the description of four-body phase space.

The description of general $n$-particle phase space in terms of
Lorentz-invariants was derived in~\cite{RevModPhys.36.595} and is
surprisingly nontrivial. The phase space is described in terms of its boundary and the phase space density. Both of these are written most conveniently by using the following formalism. For final state particles with four momenta $p_{i}^{\mu}$ one begins by introducing the matrix ${\mathcal Z}=\{z_{ij}\}$ with $z_{ij}=p_{i}\cdot p_{j}$.

Label by $\Delta_{i}$ the coefficients of the characteristic polynomial of ${\mathcal Z}$, namely the equation
\begin{eqnarray}
0&=&{\rm Det}\left[\lambda {\rm I}_{n\times n}-{\mathcal Z} \right]\nonumber\\
&=&\lambda^{n}-\left(\sum_{i=1}^{n} \Delta_{i}\lambda^{n-i}\right).
\end{eqnarray}
For example, $\Delta_{1}={\rm Tr}[{\mathcal Z}]=\sum_{i=1}^{n}m_{i}^{2}$ for any n, and $\Delta_{4}=-{\rm Det}[{\mathcal Z}]$ for $n=4$.
It is straightforward to show that this definition of $\Delta_{i}$ can also be formulated as
\begin{equation} \Delta_{l}=(-1)^{l-1}\sum{\rm determinant\ of\ all\
  }l\times l{\rm\ diagonal\ minors\ of\ }{\mathcal Z}.  \end{equation}

The kinematically accessible region for any number $n\ge4$ particles is then defined by the conditions
\begin{equation} \begin{array}{rl}\Delta_{l}>0&{\rm for\ } l\le4\\
  \Delta_{l}=0&{\rm for\ } l>4,\end{array} \end{equation} the boundary
  of the physical region corresponding to $\Delta_{4}=0$.

The $z_{ij}$ are not all independent variables. For any
given set of masses for the initial particle that decays and the final
state particles, it is straightforward to count the number of
independent phase space coordinates. The energy of each final state
particle is fixed by its three-momentum, and there are a total of $3n$
components in the three momenta of the final state particles. The
conservation of total energy and momentum for the system allows four
of these components to be expressed in terms of the rest, and finally
in situations where the decaying particle is a scalar (or its spin
states are averaged over), the rotational invariance of the system can
be used to eliminate three more degrees of freedom, leaving one with
$3n-7$ independent degrees of freedom.

The phase space density is then expressed in terms of the differential
volume element $\prod_{i<j} dz_{ij}$, and the reduction to the $3n-7$
independent degrees of freedom proceeds through the inclusion of a
number of Dirac-$\delta$ functions. While the form of these $\delta$
functions is rather complicated for
arbitrary $n$, it is very simple for $n=4$.
For all $n\ge4$, the volume element includes one $\delta$ function
that enforces total energy-momentum
conservation in terms of Lorentz-invariant variables. In the case of $n=4$, this is the only $\delta$
function present in the volume element. Indeed, there are 6 $z_{ij}$
with $i<j$, one of which can be eliminated using this $\delta$
function, leaving us with precisely 5 degrees of freedom, the correct
number for $3n-7$ when $n=4$.

Before we write down the exact form of the volume element for $n=4$, let us for future convenience switch from the set of variables $z_{ij}$, to $m_{ij}^2=(p_{i}+p_{j})^2$. Since this is a linear transformation, the functional form of the volume element is unaffected. Without further ado, the volume element of four-body phase space written in terms of Lorentz-invariant observables is
\begin{align} d\Pi_{4}
  &=\left(\prod_{i<j}dm_{ij}^{2}\right)\frac{{\rm
  C}}{M_{X}^{2}\Delta_{4}^{1/2}}\ \delta\left(\sum_{i<j}m_{ij}^{2}-{\rm
  K}\right),
  \label{eq:dLIPS4}
\end{align}
where $M_{X}$ is the mass of the original particle that decays to the four final states and
\begin{equation}
C=\frac{8}{(4\pi)^{10}}\ ,\qquad K= M_{X}^{2} + 2 \sum_{i=1}^{4} m_{i}^{2}\ .
\end{equation}

As mentioned, this volume element is not flat as in the case of the three-body phase space, but instead depends on the phase space coordinates through the factor of $\Delta_{4}^{-1/2}$ which is a function of the $m_{ij}^{2}$. Since the boundary of the kinematically allowed region is defined by $\Delta_{4}=0$,
this means that the phase space density is enhanced near the
boundary of the physical region. The apparent singularity is an
integrable one, and it is not canceled by the presence of the
$\delta$-function enforcing momentum conservation, as its argument
is linear in the $m_{ij}^{2}$ and does not lead to any nontrivial
Jacobian factors. Of course, in calculating any physical rate, the phase space density is multiplied by the square of the matrix element $|{\mathcal M}|^2$, and one may worry that on-shell intermediate particles in the cascade, in the narrow width approximation, may manifest themselves as $\delta$-functions in $|{\mathcal M}|^2$ that may change our conclusions. However, it is easily seen that the momentum going through any on-shell propagator will be a linear combination of the $m_{ij}^{2}$, thus once again these $\delta$ functions will not give rise to any Jacobians that change the functional form of the volume element, and thus cannot modify the enhancement at the boundary of the physical region.

\section{Analysis Methods}\label{sec:methods}
\begin{figure}[htp]
  \begin{center}
    \includegraphics{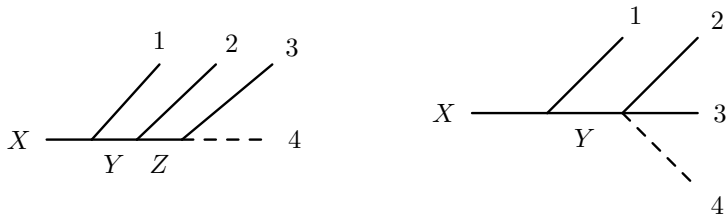}
    \qquad \qquad
    \includegraphics{topology2.eps}
  \end{center}
  \caption{The ``2+2+2'' and ``2+3'' decay topologies.}
  \label{fig:tops}
\end{figure}

In this section we will outline our analysis strategy and the setup for the
comparison with traditional endpoint methods for mass determination. Note
that there are only a few possibilities for how a cascade decay to four final
state particles may proceed in stages. Denoting the initial particle by $X$,
any intermediate resonances by $Y$, $Z$ etc., and the final state particles as 1
through 4, particle 4 always being the invisible one, the possibilities are:
\begin{itemize}
  \item $X\to 1 + Y,\  Y \to 2+Z,\ Z \to 3+4$ \\
    This is a three stage cascade decay, each stage being a two-body decay. We will refer to this possibility as the ``2+2+2'' topology.
  \item $X\to 1+Y, \ Y\to2+3+4$ \\
    This is a two stage cascade decay, with a two-body decay followed by a three-body decay, and is
    referred to as the ``2+3'' decay topology.
  \item $X\to 1+2+Y, \ Y\to 3+4$\\
    This is a two stage cascade decay, with a three-body decay followed by a two-body decay, and is
    referred to as the ``3+2'' decay topology.
  \item $X\to 1+2+3+4$\\
    A genuine four-body decay with no intermediate resonances.
\end{itemize}
The last possibility is essentially never encountered in either the SM
or any well-studied BSM scenario, so we will not dwell on it.
Furthermore, the third possibility is qualitatively very similar to
the second, as we assumed distinguishability of the final state
particles.  Thus we will not discuss the third possibility separately,
and focus our attention on the first two possibilities. These decay
topologies are shown in Figure \ref{fig:tops}. Note that we do not include in our list the ``antler'' topology, $X\to Y+Z, \ Y\to 1+2, \ Z\to 3+4$. In the literature this topology is only of interest when there are two invisible final state particles, such as particles 2 and 4 for example. Since we are concentrating on the case of a single invisible particle at the end of the cascade, one of the particles $X$ and $Y$ occur as a resonance, and should be considered as a visible final state particle in its own right. Therefore this configuration should strictly speaking be considered a three-body final state rather than a four-body one, and we omit it in our list.

Simply put, in both the phase space and the endpoint analyses, we will study
how a set of observed events in the data will favor or disfavor each one of a
large number of hypotheses for all unknown masses
$\{\tilde{m}_{\sigma}\}\equiv\{\tilde{m}_X,\tilde{m}_Y,(\tilde{m}_Z),\tilde{m}_4\}$.
In the case of the phase space analysis, this will be done by using a
likelihood, while in the endpoint analysis we will rely on a quality-of fit of
the data to any mass hypothesis that is similar to a $\chi^{2}$ variable. We
will also study how the performance of the two competing mass determination
methods will vary as the number of observed signal events changes, in
particular we will be interested in how well the two analyses perform
as one moves towards lower statistics.

It should be noted that due to the presence of two on-shell
intermediate particles, an event in the 2+2+2 case only has three
independent phase space degrees of freedom, and there are three
independent observables in the event, namely $m_{12}^{2}$, $m_{13}^{2}$
and $m_{23}^{2}$. Thus for any given mass hypothesis, an event in the
2+2+2 topology can be entirely reconstructed (or it may defy
reconstruction, thereby rejecting the mass hypothesis in question). In
fact, using information from a small number of events, the unknown
masses can be analytically solved for \cite{Hinchliffe:1998ys,
Nojiri:2003tu,
Kawagoe:2004rz,
Gjelsten:2006as,
Cheng:2007xv,
Nojiri:2008ir,
Cheng:2008mg,
Cheng:2009fw,
Matchev:2009iw,
Autermann:2009js,
Kim:2009si,
Han:2009ss,
Kang:2009sk,
Webber:2009vm,
Nojiri:2010dk,
Kang:2010hb,
Hubisz:2010ur,
Cheng:2010yy,
Gripaios:2011jm,
Han:2012nm,
Han:2012nr}. Therefore, for this
decay topology our phase space analysis is not particularly
useful over the ``polynomial method''. However, there is no similar
argument for the $2+3$ decay topology, and indeed this is the case
which will be our main focus. There
is one added subtlety in the 2+3 topology however, which is why we
choose to include a discussion of the 2+2+2 topology as a warm-up
example to this more interesting case where our methods offer genuine
improvement over existing mass determination methods.

In order to estimate the uncertainty in the results of the mass determination
procedure, we rely on a large number of pseudoexperiments. Each
pseudoexperiment is a data set with a fixed number of events generated with a
true mass point $\{M_{\sigma}\}\equiv\{M_X,M_Y,(M_Z),M_4\}$. Based on such a data
set, each mass hypothesis (distributed around the true mass point) is assigned
a quality-of-fit, which as mentioned is a likelihood in the phase space
analysis, and a $\chi^{2}$-like variable in the endpoint analysis. The mass
hypothesis with the best quality-of-fit for a given data set is then chosen as
the ``winner'' for that pseudoexperiment. As we repeat this procedure over a
large number of pseudoexperiments, a distribution of winning hypotheses
emerges. We use the statistical properties of this distribution of winners in
order to determine both the central value and the uncertainty of the mass
determination procedure.

\subsection{Warming Up: The 2+2+2 topology}

\subsubsection{Analysis using four-body phase space}

We set up the likelihood analysis based on the phase space volume element as
follows: For an event with kinematic observables $m_{12}^{2}, m_{13}^{2}$ and
$m_{23}^{2}$ we first use the mass hypothesis $\{\tilde{m}_{\sigma}\}$ and
overall energy momentum conservation to solve for $m_{i4}^2$. Note that unless
the hypothesis used is the true mass point, these will not be the true values
of $m_{i4}^2$. We then define the likelihood as the normalized probability of
obtaining this event given the mass hypothesis $\{\tilde{m}_{\sigma}\}$
\begin{equation}
  \mathcal{L}(\tilde{m}_\sigma, m_{ij}^{2})=
  \frac{1}{\Gamma_X}
  \frac{1}{2 \tilde{m}_X}
  \mu^2_{XY1} \mu^2_{YZ2} \mu^2_{Z34}
  \frac{\pi}{\tilde{m}_Y \Gamma_Y}
  \frac{\pi}{\tilde{m}_Z \Gamma_Z}
  \frac{1}{(4\pi)^6 2\tilde{m}_X^2}
  \Theta(\Delta_4)
  \frac{1}{\Delta_{4}^{1/2}}.
\end{equation}
Several comments are in order. All widths $\Gamma$ as well as $\Delta_4$ appearing in this formula are computed with respect to the mass hypothesis $\{\tilde{m}_{\sigma}\}$. The first factor of $\frac{1}{\Gamma_X}$ simply ensures that the likelihood is normalized to one for any mass hypothesis
\begin{equation}
  \int
  \mathcal{L}(\tilde{m}_\sigma, m_{ij}^{2})
  \ dm_{12}^{2} dm_{13}^{2} dm_{23}^{2}=1.
\end{equation}
The factor of $\frac{1}{2 \tilde{m}_X}$ is simply the prefactor in the
definition of a differential width. The trilinear couplings $\mu$ have been
labeled in the obvious notation, and have mass dimension one. The factors of
$\frac{\pi}{\tilde{m}_Y \Gamma_Y}$ and $\frac{\pi}{\tilde{m}_Z \Gamma_Z}$ are
simply the remnants of the $Y$ and $Z$ propagators after the narrow width
approximation,
\begin{align}
  \frac{1}{
  |\tilde{m}_Y^2
  -
  (p_2+p_3+p_4)^2
  + i\, \tilde{m}_Y \Gamma_Y |^2}
  &\simeq
  \delta(
  m_{23}^{2}+m_{24}^{2}+m_{34}^{2}- K_Y)
  \frac{\pi}{\tilde{m}_Y \Gamma_Y},
\nonumber \\ & \qquad K_Y = \tilde{m}_Y^2 + m_2^2 + m_3^2 + m_4^2,
  \\
  \frac{1}{
  |\tilde{m}_Z^2
  -
  (p_3+p_4)^2
  + i\, \tilde{m}_Z \Gamma_Z |^2}
  &\simeq
  \delta(
  m_{34}^{2}-\tilde{m}_Z^2)
  \frac{\pi}{\tilde{m}_Z \Gamma_Z}
  \ .
  \label{eq:YZOnShell}
\end{align}
The factors of $\mu$ and these propagator remnants simply stand for the squared matrix element $|{\mathcal M}|^2$ when all involved particles are scalars. The remaining factors simply correspond to the phase space volume element given in equation~\ref{eq:dLIPS4}. The step function $\Theta(\Delta_4)$ appears because for an incorrect mass hypothesis, a given event may simply not reconstruct a physical solution, in which case the likelihood evaluates to zero. Note that when finite detector resolution effects are taken into
account, the step function needs to be convoluted with the energy resolution of
the detector, leading to a smoother turn-off when an event proves difficult to
reconstruct. We will come back to this issue in section \ref{sec:conclusions}.

It is easily seen that there are several cancellations in the likelihood formula between the factors of $\Gamma$ and $\mu^{2}$, leaving only dimensionless functions depending on the masses. In particular, using the respective two-body decay formulae for the widths of $X,Y,Z$,
we obtain the final expression for the likelihood
\begin{align}
  \mathcal{L}(\tilde{m}_\sigma,m_{ij}^{2})
  &\simeq
  \frac{1}{4\pi \tilde{m}_X^2}
  \left(
  1-\frac{\tilde{m}_Y^2}{\tilde{m}_X^2}
  \right)^{-1}
  \left(
  1-\frac{\tilde{m}_Z^2}{\tilde{m}_Y^2}
  \right)^{-1}
  \left(
  1-\frac{\tilde{m}_4^2}{\tilde{m}_Z^2}
  \right)^{-1}
  \Theta(\Delta_4)
  \frac{1}{\Delta_{4}^{1/2}}\,
\end{align}
 where we have chosen a simplified form for the dimensionless phase
space functions by assuming $m_{1,2,3} \simeq 0$. We use this simpler
expression in our analysis, however we do verify numerically that for
$m_{1,2,3}\simeq$ few GeV, the difference with the exact result is negligible.
The likelihood for the pseudoexperiment is simply the product of likelihoods of each event,
\begin{align}
  \mathcal{L}(\tilde{m}_\sigma)
  &=
  \prod_{events}
  \mathcal{L}(\tilde{m}_\sigma,m_{ij}^{2}).
\end{align}

For each pseudoexperiment, the winning mass hypothesis is then chosen as the one that maximizes the above likelihood. In practice we use the log-likeliood and sum over the events in the data sample.

\subsubsection{Using edges and endpoints}

We want to compare the results of this phase space analysis
with those obtained from the endpoint analysis. As noted
before, this is not the most optimal analysis for the 2+2+2
topology as the ``polynomial method'' lets us reconstruct the masses
analytically. However, the comparison we set up in this topology will prepare us for the 2+3 topology
considered in the next subsection.

The mass determination procedure of the endpoint analysis is based on
comparing for each pseudoexperiment the measured position of kinematic features
obtained from the data with those that are predicted based on the mass
hypothesis being used. In appendix \ref{sec:edges-endpoint-form}, we list all
the formulae the predicted positions of the
kinematic edges and endpoints for this topology using a given mass hypothesis.
For the 2+2+2 topology, $Y$ and $Z$ are on-shell, therefore the $m_{12}^{2}$
and $m_{23}^{2}$ distributions have an edge. In contrast, the $m_{13}^{2}$ and
$m_{123}^{2}$ distributions have endpoints. The measured positions of the same
kinematic edges and endpoints are defined simply as the highest value obtained
for the observable in question using all the events in the data set.

We then define the quality-of-fit variable for any mass hypothesis
based on a pseudoexperiment as
\begin{align}
  Q
  &=
  \left(\sum_{i={\rm endpts.}}
  \left(\frac{\mathcal{O}_{i,predicted} - \mathcal{O}_{i,measured}}
{\mathcal{O}_{i,measured}}\right)^2\right)
\mathcal{F}
\, ,
  \label{eq:qualityoffit}
\end{align}
where the observables $\mathcal{O}_{i} = \{(m_{123}^2)_{max},
(m_{12}^2)_{max}, (m_{23}^2)_{max}, (m_{13}^2)_{max} \}$ represent the
predicted and measured positions of each edge and endpoint, and the
sum is taken over all relevant edges and endpoints.  $\mathcal{F}=1$
if all measured endpoints are consistent with the predicted ones,
meaning that they occur at smaller (or equal) values of
$\mathcal{O}_i$. If any one of the measured endpoints exceeds the
predicted value, the mass hypothesis is rejected (equivalently,
$\mathcal{F}$ is taken to be $\infty$ in that case).  As we remarked
for the phase space analysis, this sharp step-function behavior will
need to be softened once a more realistic model of the detector with
finite resolution is used. It should also be remarked here that while
it may appear somewhat arbitrary to assign the same weight in $Q$ to
the contribution of all edges and endpoints, practically $Q$ is almost
always dominated by the position of one of the edges or endpoints,
thus the weighting has no significant impact. We dwell on this point
in more detail in appendix \ref{sec:edge-end-point}.

\subsection{Focus Case: The 2+3 topology}

We now consider the case where $Y$ has a 3-body decay to $2,3,4$ without an
intermediate resonance.  The added complication compared to the
2+2+2 topology is that since we have lost the on-shell constraint of $Z$, this
event topology
is now impossible to reconstruct analytically. In other words, while there are
still three observables $m_{ij}^{2}$ as in the 2+2+2 topology, there are now
four independent degrees of freedom in the phase space, since only the $Y$
propagator can be used in the narrow width approximation to eliminate one of
the phase space integration variables.

\subsubsection{Analysis using four-body phase space}

One unmeasured phase space variable, which we take to be $m^2_{34}$, can no longer be fixed
given a mass hypothesis. Hence we integrate over this parameter in the definition of the likelihood.
Following the same procedure as earlier, we can use momentum
conservation and the on-shell $Y$ propagator to define the likelihood
\begin{equation}
  \mathcal{L}(\tilde{m}_\sigma, m_{ij}^{2})=
  \frac{1}{\Gamma_X}
  \int dm_{34}^{2}
  \frac{1}{2 \tilde{m}_X}
  \mu^2_{XY1} \lambda^2_{Y234}
  \frac{\pi}{\tilde{m}_Y \Gamma_Y}
  \frac{1}{(4\pi)^6 2\tilde{m}_X^2}
  \Theta(\Delta_4)
  \frac{1}{\Delta_{4}^{1/2}}\ .
  \label{eq:23likelihoodA}
\end{equation}
Note that $Y$ decays through a quartic scalar coupling which we have labeled as
$\lambda$ since it is dimensionless. The factor of $\Theta(\Delta_4)$ now
determines the allowed range of $m_{34}^{2}$ integration, and for an incorrect
mass hypothesis it is possible that no value of $m_{34}^{2}$ will lead
to a possible reconstruction of the event. As in the 2+2+2 topology, in that case
the mass hypothesis may be ruled out with a single event.

The $m_{34}^{2}$ integration can be performed analytically. We present
the details of this calculation in appendix \ref{sec:integrating-m34} and
simply list the result here, which is
\begin{align}
  \mathcal{L}(\tilde{m}_\sigma, m_{ij}^{2})
  &=
  \frac{1}{\Gamma_X}
  \frac{1}{2 \tilde{m}_X}
  \mu^2_{XY1} \lambda^2_{Y234}
  \frac{\pi}{\tilde{m}_Y \Gamma_Y}
  \frac{1}{(4\pi)^5 2\tilde{m}_X^2}
  \frac{1}{\sqrt{\lambda_0}}
  \Theta(-G_1) \Theta(-G_2)\,,
  \label{eq:23likelihoodB}
\end{align}
where
\begin{align}
  \lambda_0 &= \lambda\left(m_1^2, m_{23}^2, m_{123}^2\right), \\
  G_1 &= G\left(m_{12}^2, m_{23}^2, m_{123}^2, m_2^2, m_1^2, m_3^2\right),\\
  G_2 &= G\left(m_{123}^2, \tilde{m}_{Y}^2, \tilde{m}_{X}^2, m_{23}^2, m_{1}^2, \tilde{m}_4^2\right),
\end{align}
and the kinematic functions $\lambda$ and $G$ are defined as~\cite{BycklingKajantie197301}.
\begin{align}
  \lambda(X,Y,Z) &= X^2 + Y^2 + Z^2 - 2 XY - 2YZ - 2 ZX,\nonumber \\
  G(X,Y,Z,U,V,W)
  &= X Y (X+Y-Z-U-V-W)+Z U (Z+U-X-Y-V-W)
  \nonumber \\&
  +V W (V+W-X-Y-Z-U)+XZW +XUV +YZV +YUW.
  \label{eq:lambdaG}
\end{align}
The likelihood is normalized to 1 as in the 2+2+2 case.

Again using a somewhat more simplified expression assuming $m_{1,2,3} \simeq 0$, and canceling factors of $\Gamma$ against $\mu^{2}$ and $\lambda^{2}$, we can rewrite this as
\begin{align}
  \mathcal{L}(\tilde{m}_\sigma, m_{ij}^{2})
  &\simeq
  \frac{1}{2\tilde{m}_X^2}
  \left(
  1-\frac{\tilde{m}_Y^2}{\tilde{m}_X^2}
  \right)^{-1}
  \frac{1}{\tilde{m}_Y^2}
  \left(1 -
  \frac{\tilde{m}_4^4}{\tilde{m}_Y^4}
  -2\frac{\tilde{m}_4^2}{\tilde{m}_Y^2}
  \log\left(\frac{\tilde{m}_Y^2}{\tilde{m}_4^2}\right)\right)^{-1} \times
\nonumber \\ &
\qquad \qquad \qquad
\qquad \qquad \qquad
\qquad \qquad \qquad
  \Theta(-G1) \Theta(-G2)
  \frac{1}{\sqrt{\lambda_0}} .
\end{align}

As before, the likelihood for
the pseudoexperiment is simply the product of the likelihoods for each
event,
\begin{align}
  \mathcal{L}(\tilde{m}_\sigma)
  &=
  \prod_{events}
  \mathcal{L}(\tilde{m}_\sigma,m_{ij}^{2}).
\end{align}

\subsubsection{Using edges and endpoints}

Unlike the case of the phase space analysis, the endpoint
analysis for the 2+3 topology is not significantly different than in the 2+2+2
topology. The absence of $Z$ as an on-shell state demotes the edges in the
$m_{12}^{2}$ and $m_{23}^{2}$ distributions to
endpoints, but the quality-of-fit variable $Q$ is defined in the same way as in
equation \ref{eq:qualityoffit}. Since our focus is on situations with low
statistics, this often means that the endpoint is not well populated, leading
to a large underestimation of the measured endpoint position. Therefore, we
expect the
results in this case to be less accurate than those for the 2+2+2 topology.

We discuss the results for all analyses described above in detail in the next section using a specific choice of true mass point $\{M_{\sigma}\}$.

\section{Results} \label{sec:results}
After having outlined the theoretical details of the analysis we wish to set
up, we now turn to the practical implementation and report our results in this
section.

The data samples on which we run both the endpoint and the
phase space analyses are generated at the parton level using MadGraph 5
\cite{Alwall:2011uj}. As mentioned in the introduction, in order to separate
the effects of spin-correlations from the effects of the phase space density,
we choose all particles as scalars in the MadGraph model, and we choose one
specific mass spectra to use as benchmarks in the analysis of each one of the two decay
topologies. All analyses rely on parton-level events as we have chosen to work
with a detector with perfect resolution.

\subsection{The 2+2+2 Topology}

We work with the benchmark spectrum
\begin{align}
  \nonumber
  M_X &= 500 \textrm{ GeV}\,, \qquad
  M_Y = 350 \textrm{ GeV}\,, \\
  \nonumber
  M_Z &= 200 \textrm{ GeV}\,, \qquad
  M_4 = 100 \textrm{ GeV}\,, \\
  &m_1 = m_2 = m_3 = 5   \textrm{ GeV}\,,
  \label{eq:truemass222}
\end{align}
and for each pseudoexperiment we use data samples with  $N_{events} =
100$.

In both the phase space and the endpoint analyses, we also need a
set of mass hypotheses $\{\tilde{m}_{\sigma}\}$ from which one is
chosen as the winner of each pseudoexperiment. In order to get an
accurate picture of the uncertainty in the mass measurement, we check that
the range of the set of
mass hypotheses is large enough such that a further enlargement would
have no effect. In
particular, we verify that the flat direction does not
extend past the boundaries of the chosen set of mass hypotheses. Below
we list how we choose the set of mass hypotheses for both analysis
methods, and we later confirm as we list the results of our analysis
that the chosen sets indeed satisfy this criterion.

We
choose a set of mass hypotheses for the
analysis such that the expected flat direction, where all hypothesis masses
are raised or lowered together, is well populated and
finely scanned. Thus, instead of generating a set of mass hypotheses
that lie on a (hyper)cubic lattice in the
$\{\tilde{m}_X,\tilde{m}_Y,\tilde{m}_Z,\tilde{m}_4\}$ basis, we
generate a lattice along a different set of axes, namely:
\begin{equation}
\tilde{m}_{\sigma}=M_{\sigma}+(100~{\rm GeV})
\left(\alpha V_{\sigma}^{(1)}+\beta V_{\sigma}^{(2)}+\gamma V_{\sigma}^{(3)}+\delta V_{\sigma}^{(4)}\right) \qquad \sigma=\{X,Y,Z,4\}\,,
\end{equation}
where
\begin{eqnarray}
V_{\sigma}^{(1)}&=&\{1,1,1,1\}, \nonumber \\
V_{\sigma}^{(2)}&=&\{1,-1,0,0\}, \nonumber \\
V_{\sigma}^{(3)}&=&\{1,1,-1,-1\}, \nonumber\\
V_{\sigma}^{(4)}&=&\{0,0,1,-1\} \,.
\end{eqnarray}
Note that these vectors are chosen to give an orthogonal basis.
$V_{\sigma}^{(1)}$ is the vector that corresponds to the flat
direction.
For the endpoint analysis, we use the following scan for $\alpha,\beta, \gamma,
\delta$, \begin{align}
\alpha &\in [-1,7],
\nonumber \\
\beta,\gamma,\delta &\in [-0.3,0.3],
\nonumber \\
 \Delta \alpha &= \Delta \beta = \Delta \gamma = \Delta \delta = 0.02
 \,.
\end{align}
Any choice of
$(\alpha,\beta,\gamma,\delta)$ is discarded for which the inequality
 $\tilde{m}_X>\tilde{m}_Y>\tilde{m}_Z>\tilde{m}_4>0$
is violated. We have chosen to scan a much wider interval for $\alpha$
in order to accurately sample the flat direction.

For the phase space analysis, we provide results from a much finer scan
in a narrow range (after verifying that all winning hypotheses indeed
lie in this narrow range). For this case we choose
\begin{align}
\alpha &\in
[-1.5\times10^{-2}, 1.5\times10^{-2}] & \Delta \alpha &=
10^{-3},
\nonumber \\
\beta,\gamma,\delta &\in [-1.5\times 10^{-3},1.5\times 10^{-3}] &
 \Delta \beta &= \Delta \gamma = \Delta \delta = 10^{-4}.
\end{align}

\begin{table}[tp]
  \begin{tabular}{|c|c|c|}
    \hline\hline
    Mass (GeV) &\quad Phase space \quad &\qquad  End-points \quad \qquad \\
    \hline
    $m_{X}$ & $499.89 \pm 0.60$ & $677.41 \pm 157.47$\\
    $m_{Y}$ & $349.90 \pm 0.59$ & $527.19 \pm 155.96$\\
    $m_{Z}$ & $199.92 \pm 0.59$ & $380.11 \pm 160.57$\\
    $m_{4}$ & $ 99.93 \pm 0.65$ & $277.87 \pm 156.42$\\
    \hline
    $\alpha$ & $(-0.87 \pm 6.03) \times 10^{-3}$ & $1.78 \pm 1.58 $\\
    $\beta$ & $(-0.07 \pm 0.38) \times 10^{-3}$ & $(0.11 \pm 1.54)
    \times 10^{-2}$\\
    $\gamma$ & $(-0.17 \pm 0.44) \times 10^{-3}$ & $(-0.84 \pm 1.44)
    \times 10^{-2} $\\
    $\delta$ & $(-0.09 \pm 0.66) \times 10^{-3}$ & $(1.12 \pm 3.08)
    \times 10^{-2}$\\
    \hline\hline
  \end{tabular}
  \caption{
  Mean and standard deviation of the distribution of winning mass
  hypotheses across pseudoexperiments using the phase space and
  endpoint analyses for the 2+2+2 topology. The number of events in
  each pseudoexperiment was chosen to be $N_{events} = 100$.
  }
  \label{tab:222}
\end{table}

\begin{figure}[tp]
  \begin{center}
    \includegraphics[width=0.45\textwidth]{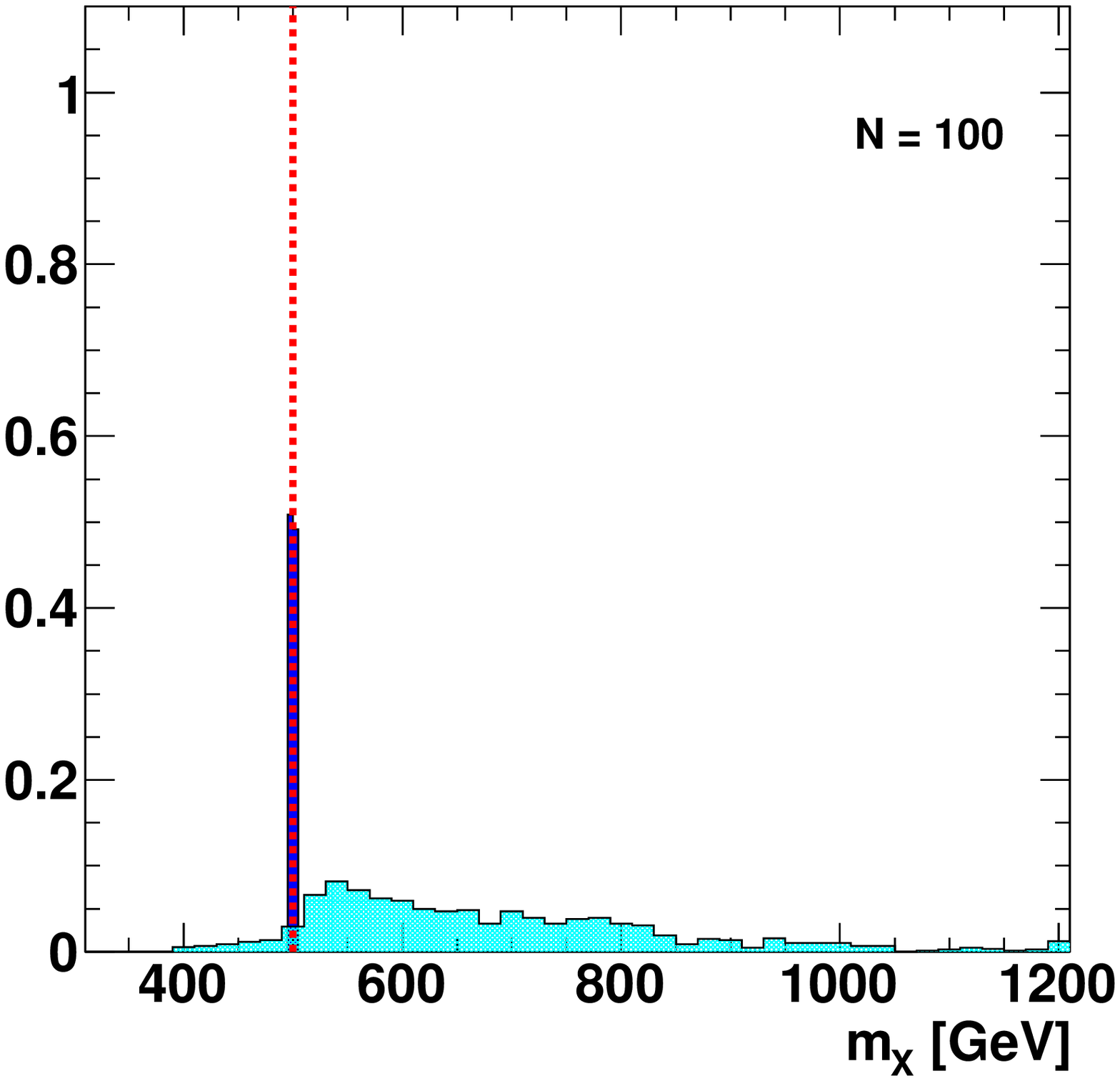}\qquad
    \includegraphics[width=0.45\textwidth]{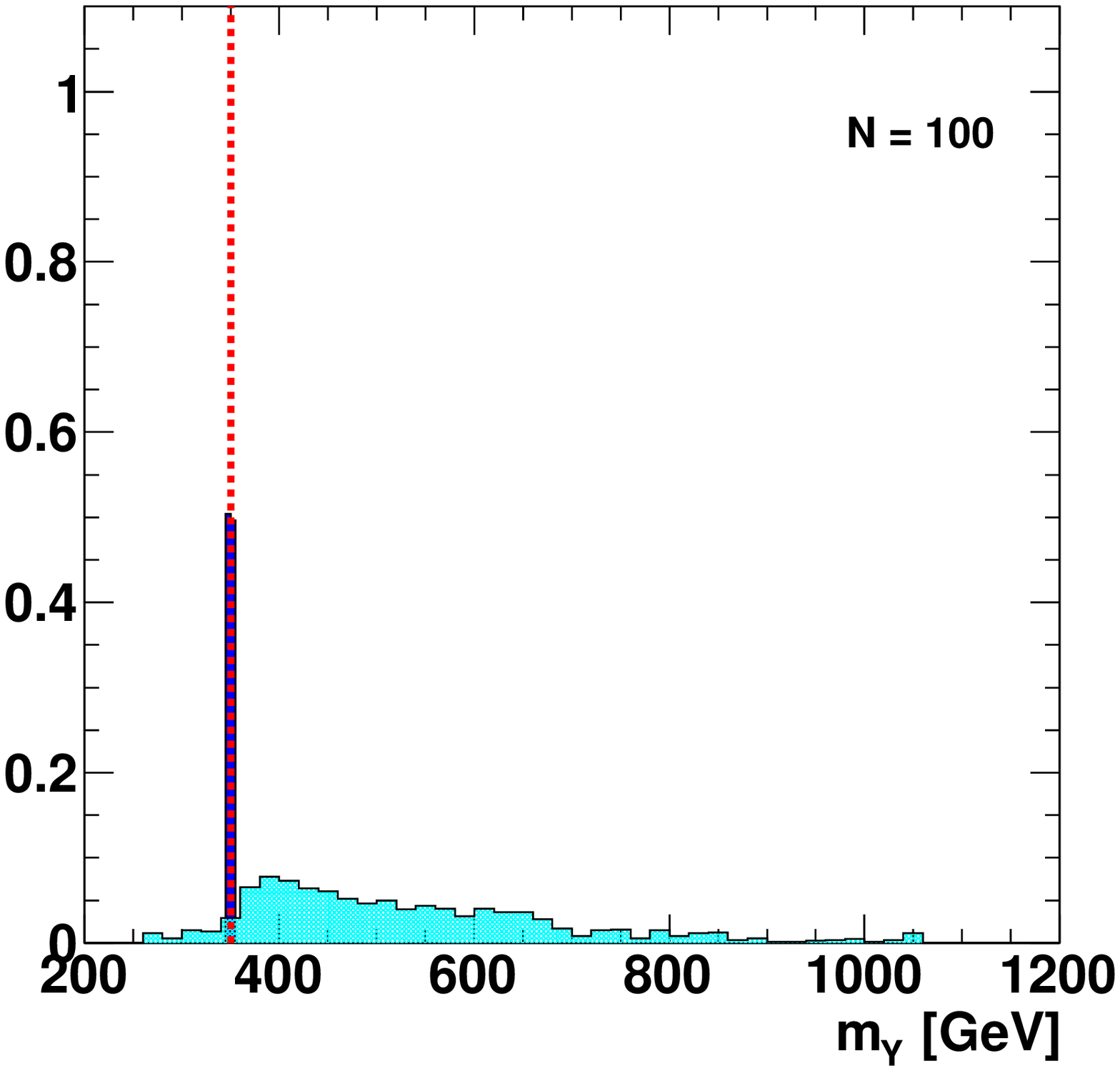}\\
    \includegraphics[width=0.45\textwidth]{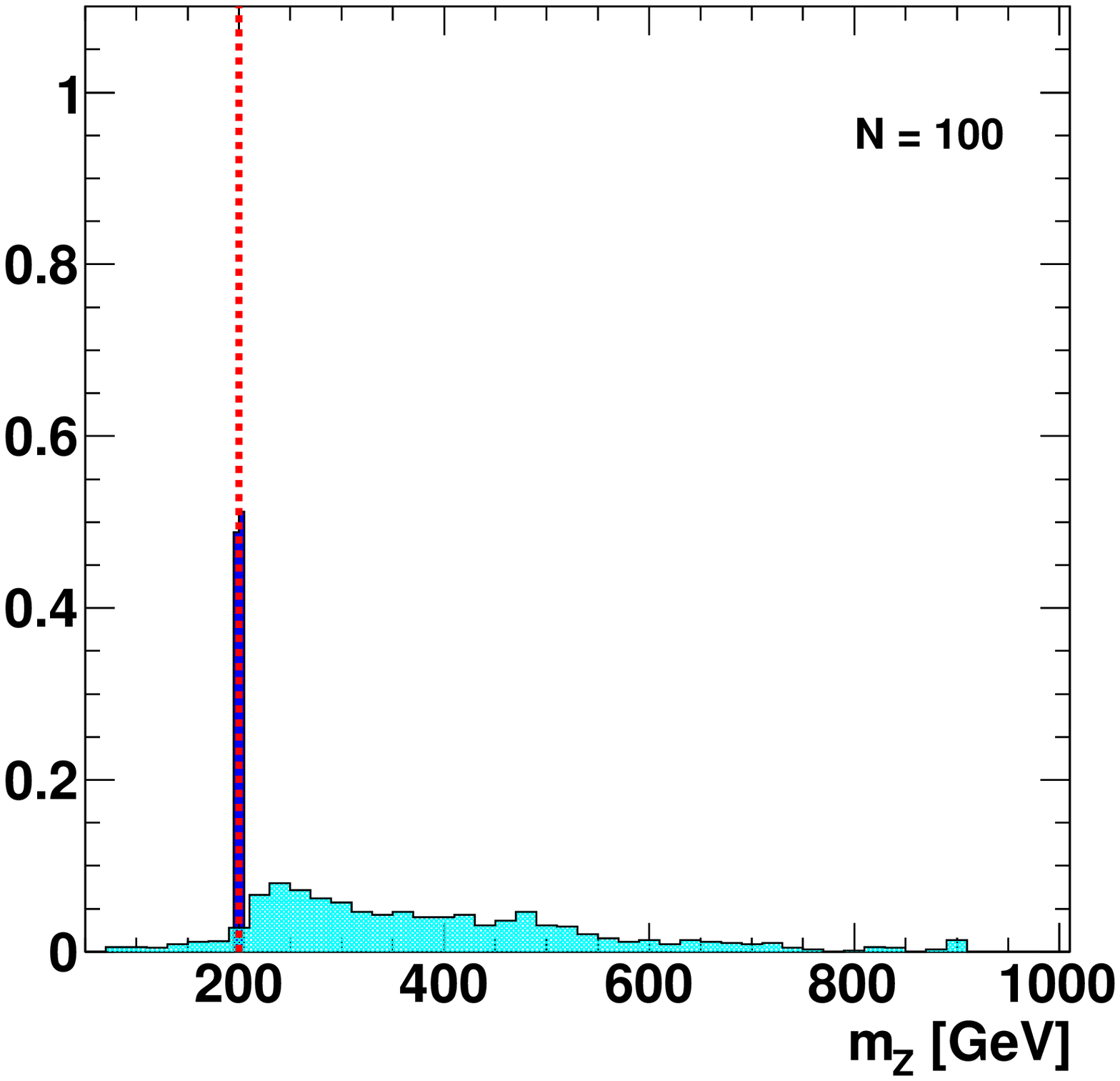}\qquad
    \includegraphics[width=0.45\textwidth]{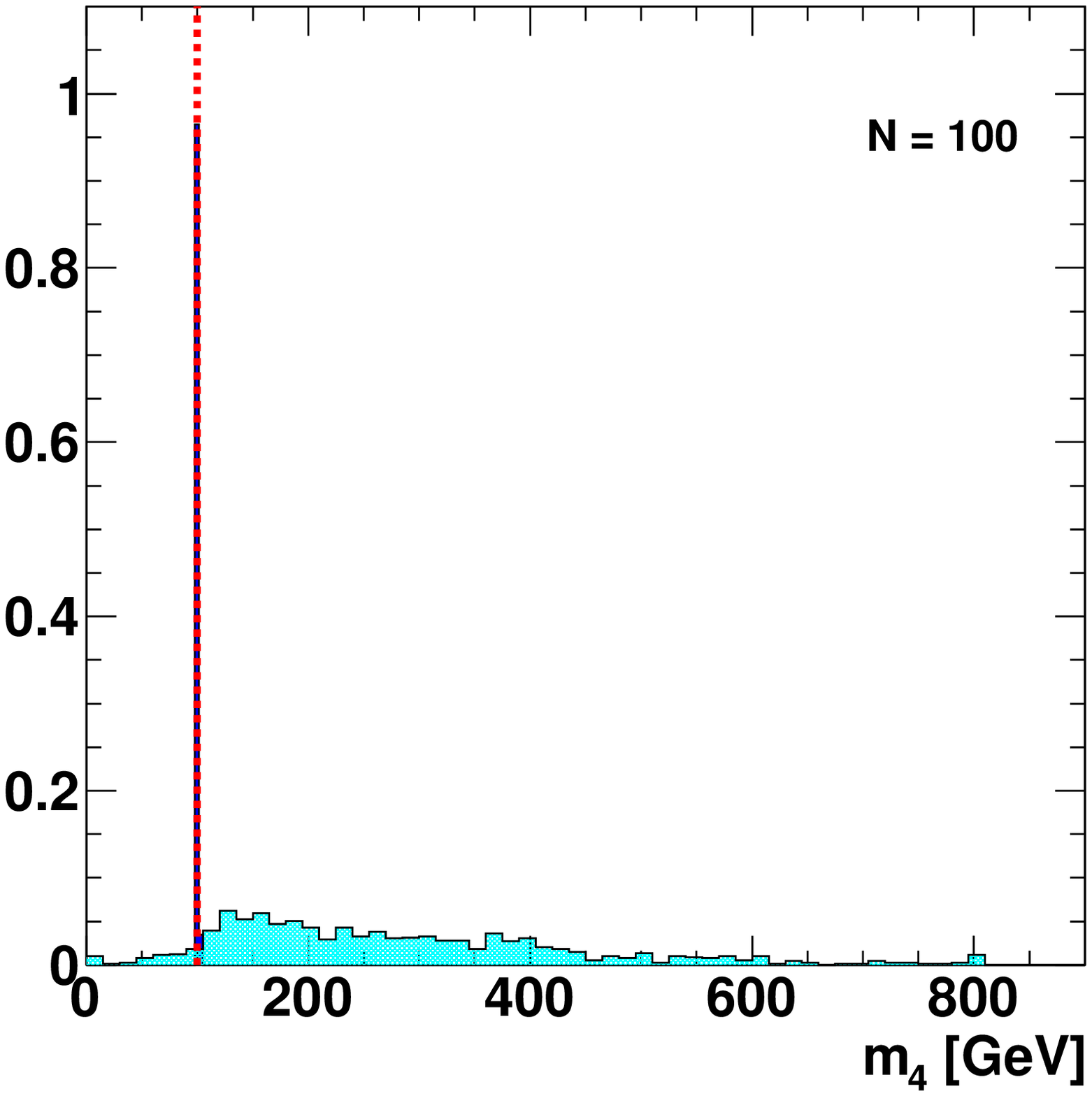}
    \caption{
    The distribution of winning mass hypotheses across pseudoexperiments
    using the phase space analysis (blue) and endpoint analysis
    (cyan) for the
    ``2+2+2'' topology. The true mass values in the benchmark spectrum used to
    generate the data are shown by dashed red lines.}
    \label{fig:222X}
  \end{center}
\end{figure}

\begin{figure}[tp]
  \begin{center}
    \includegraphics[width=0.45\textwidth]{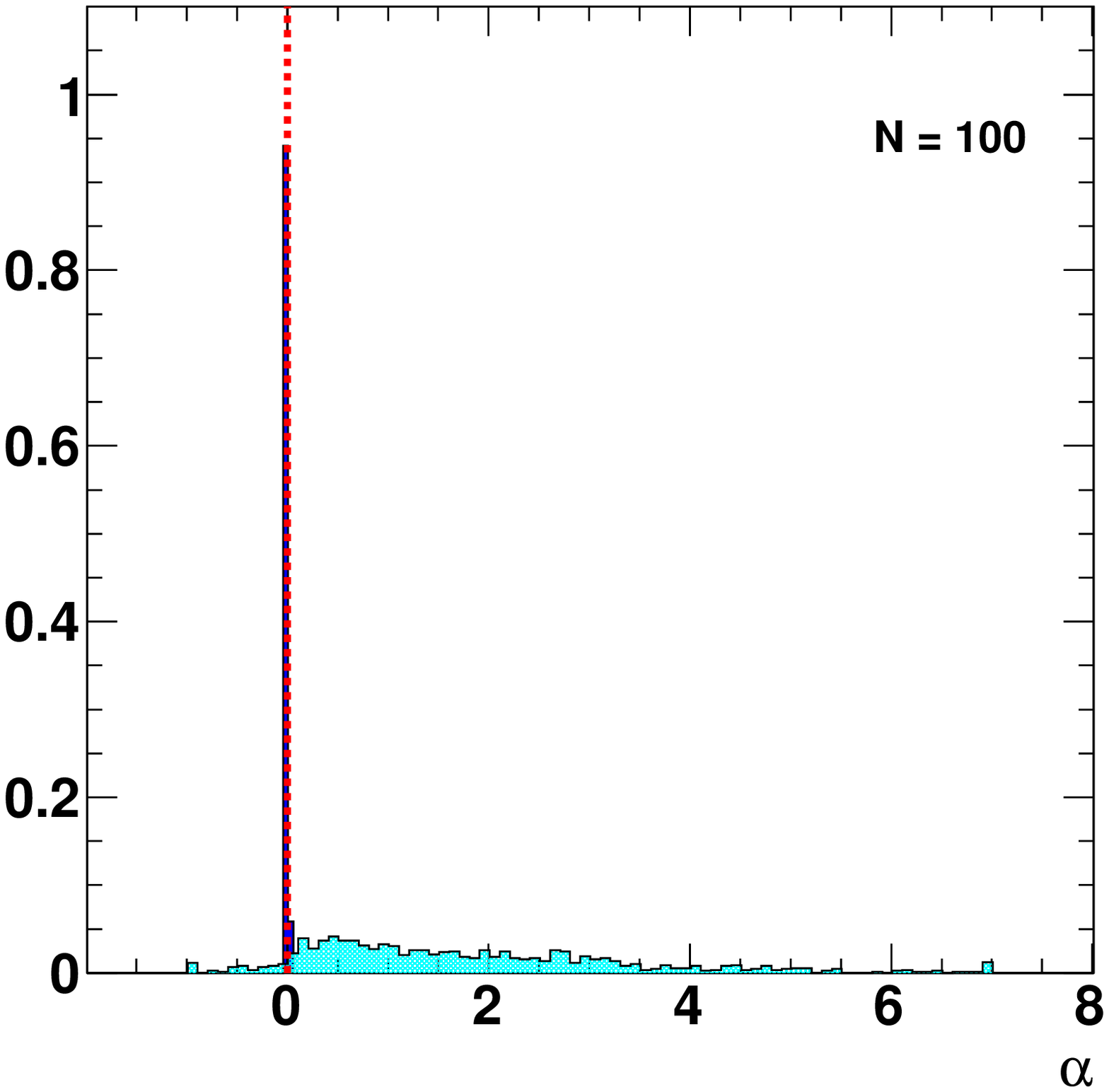}\qquad
    \includegraphics[width=0.45\textwidth]{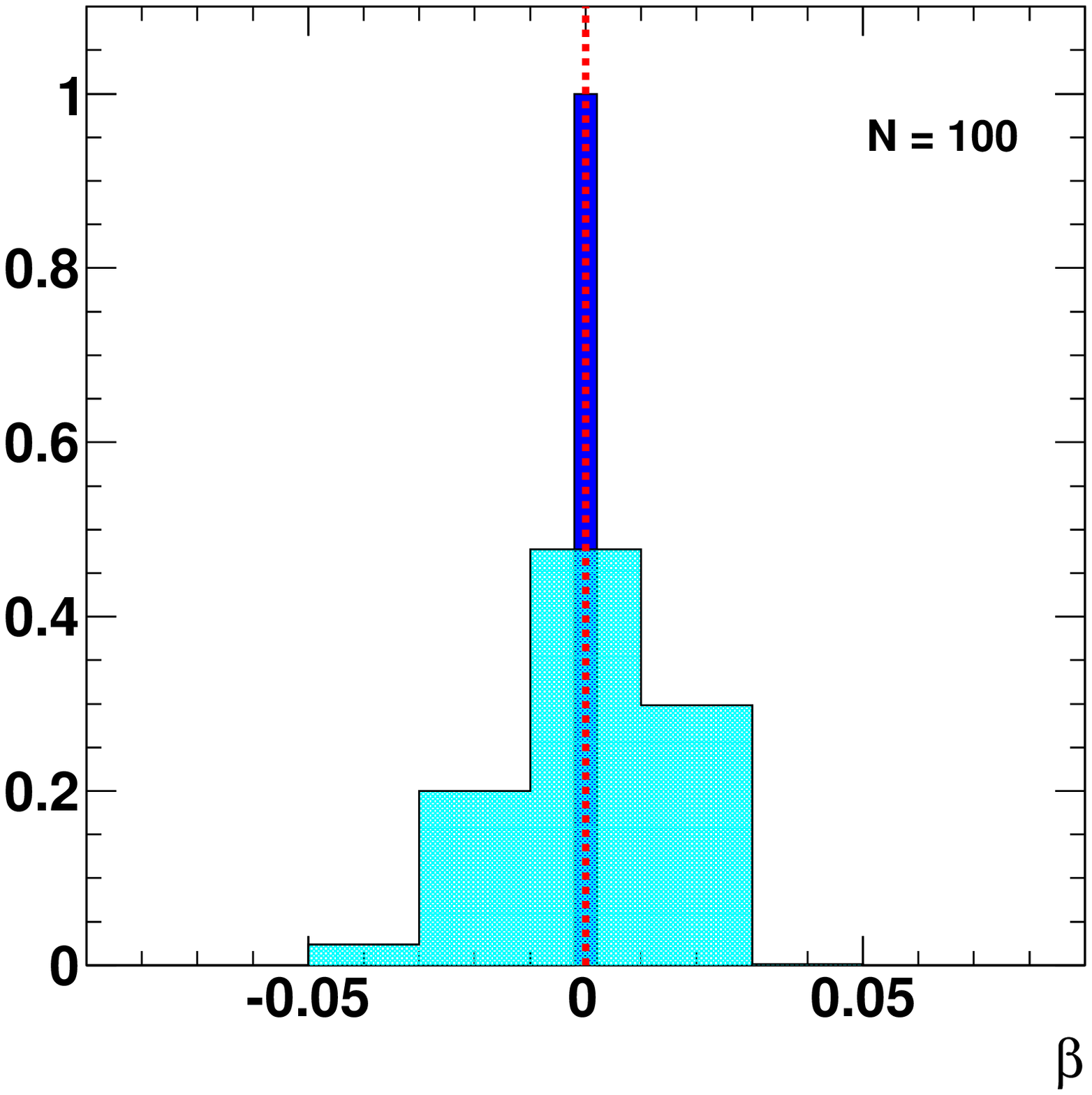}\\
    \includegraphics[width=0.45\textwidth]{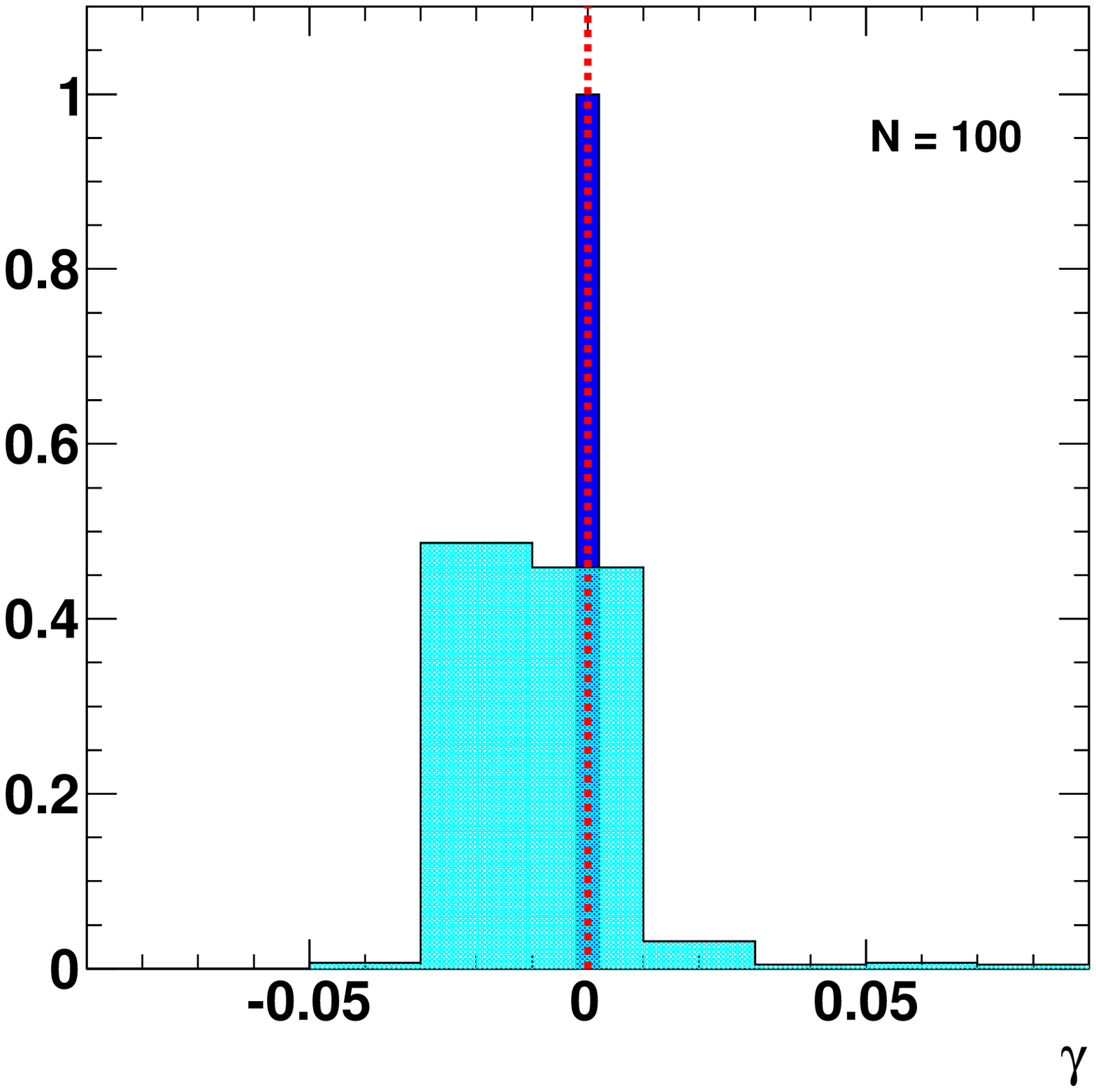}\qquad
    \includegraphics[width=0.45\textwidth]{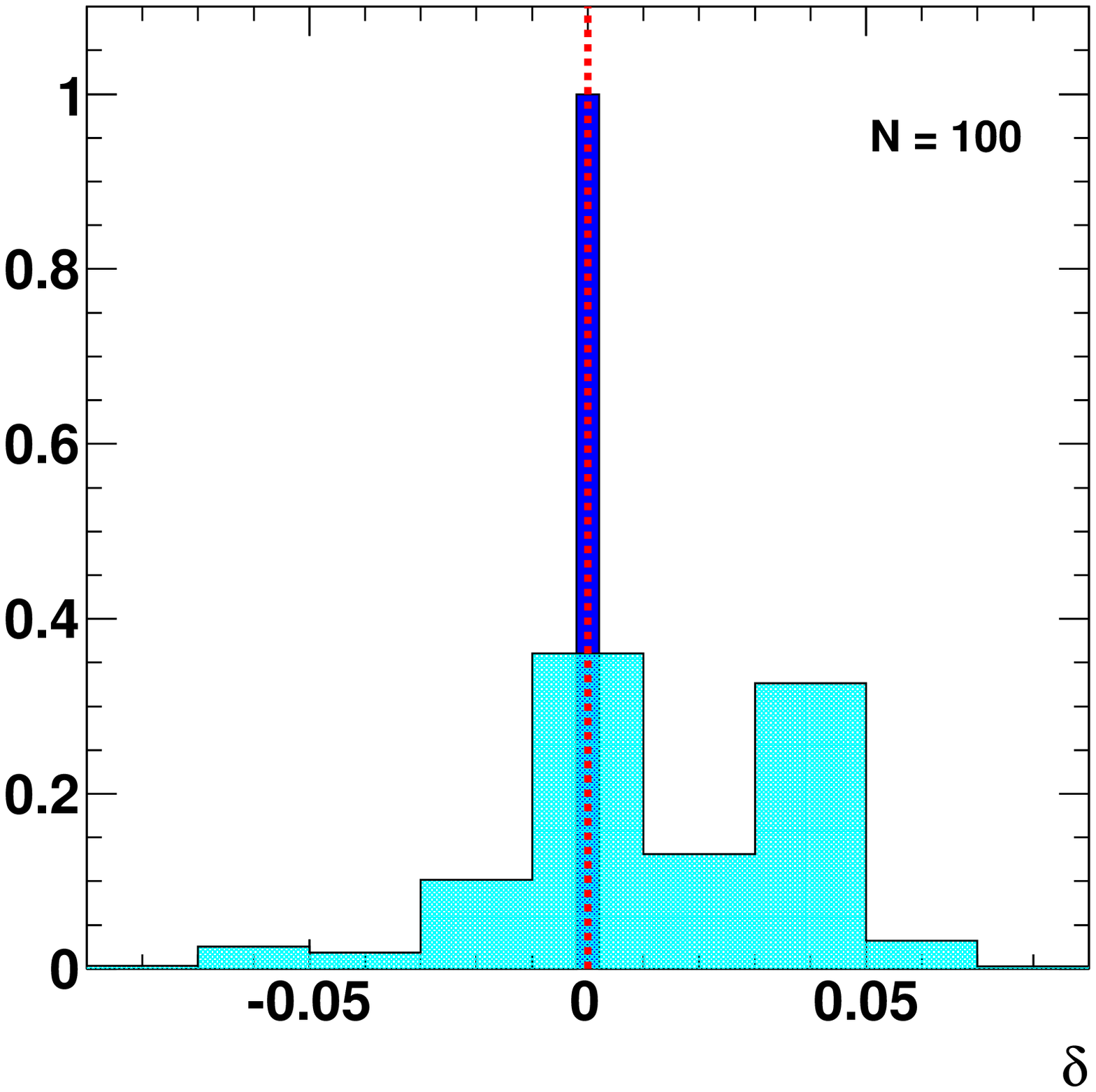}
    \caption{
    The distribution of winning mass hypotheses across pseudoexperiments
    using the phase space analysis (blue) and endpoint analysis
    (cyan) for the
    ``2+2+2'' topology in terms of $\{\alpha, \beta, \gamma, \delta \}$. The
true mass values in the benchmark spectrum used to generate the data are shown
by dashed red lines.}
    \label{fig:222a}
  \end{center}
\end{figure}

Having chosen a benchmark spectrum and a set of mass hypotheses, we
proceed to run both the phase space and endpoint analyses on
${\mathcal O}(10^3)$ pseudoexperiments. The winner of each
 pseudoexperiment is chosen in the two analyses as explained in detail
in the previous section. In figure \ref{fig:222X} and \ref{fig:222a} we plot the
distribution of winning mass hypotheses in the two analyses.
We also list the mean and standard deviation in the
distribution of these quantities in table~\ref{tab:222}. The difference
in sensitivity between the two analysis methods is dramatic. Perhaps
most importantly, one can observe that the phase space analysis based
on the four-body phase space has much better resolution along the flat
direction that causes the endpoint analysis results to acquire an
enormous spread and bias.

We should note that the standard deviations of the distribution of
winners in the $(\alpha,\beta,\gamma,\delta)$ parametrization give a
clearer indication of the inherent uncertainty in the mass
determination between these two methods. The standard deviations in
the distribution of winners in the
$\{\tilde{m}_X,\tilde{m}_Y,\tilde{m}_Z,\tilde{m}_4\}$ parametrization
on the other hand should not be regarded as independent measurement
uncertainties since they are all dominated by the spread along the
flat direction scanned by $\alpha$.

In appendix \ref{sec:edge-end-point} we investigate the endpoint analysis
along the flat direction in further detail, and show why it
produces a poor result at low statistics, and furthermore
exhibits a bias towards higher mass values.

\subsection{The 2+3 Topology}

For the 2+3 topology we choose to work with the benchmark spectrum
\begin{align}
  \nonumber
  m_X &= 500 \textrm{ GeV}\,, \qquad
  m_Y = 350 \textrm{ GeV}\,, \qquad
  m_4 = 100 \textrm{ GeV}\,, \\
  &\qquad\qquad m_1 = m_2 = m_3 = 5   \textrm{ GeV}\,.
\end{align}
which follows from the benchmark spectrum we used for the 2+2+2 topology by making $Z$
very heavy, so that it is always off-shell.

We use the following set of mass hypotheses for both the phase space
and the endpoint analyses:
\begin{equation}
\tilde{m}_{\sigma}=M_{\sigma}+(100~{\rm GeV})\left(\alpha V_{\sigma}^{(1)}+\beta V_{\sigma}^{(2)}+\gamma V_{\sigma}^{(3)}\right) \qquad \sigma=\{X,Y,4\}\,,
\end{equation}
where
\begin{align}
V_{\sigma}^{(1)}&=\{1,1,1\} \nonumber \\
V_{\sigma}^{(2)}&=\{0,1,-1\} \nonumber \\
V_{\sigma}^{(3)}&=\{2,-1,-1\}\,.
\end{align}
Once again, $V_{\sigma}^{(1)}$ corresponds to the flat direction. For
the phase space analysis, we choose the scan
\begin{align}
\alpha &\in [-1,1] & \Delta \alpha &=0.01,  \nonumber \\
\beta &\in [-0.1,0.1] & \Delta \beta &= 0.005,  \nonumber \\
\gamma &\in [-0.05,0.05] & \Delta \gamma &= 0.0025 \,.
\end{align}
As before, choices
of $(\alpha,\beta,\gamma)$ are discarded for which the inequality
 $\tilde{m}_X>\tilde{m}_Y>\tilde{m}_4>0$ is violated.
For the endpoint analysis, we scale this grid up by a factor of 2,
since in this case a wider range of mass hypotheses can be picked as
the winners.

\begin{figure}[tp]
  \begin{center}
    \subfloat[]{
    \includegraphics[scale=0.25]{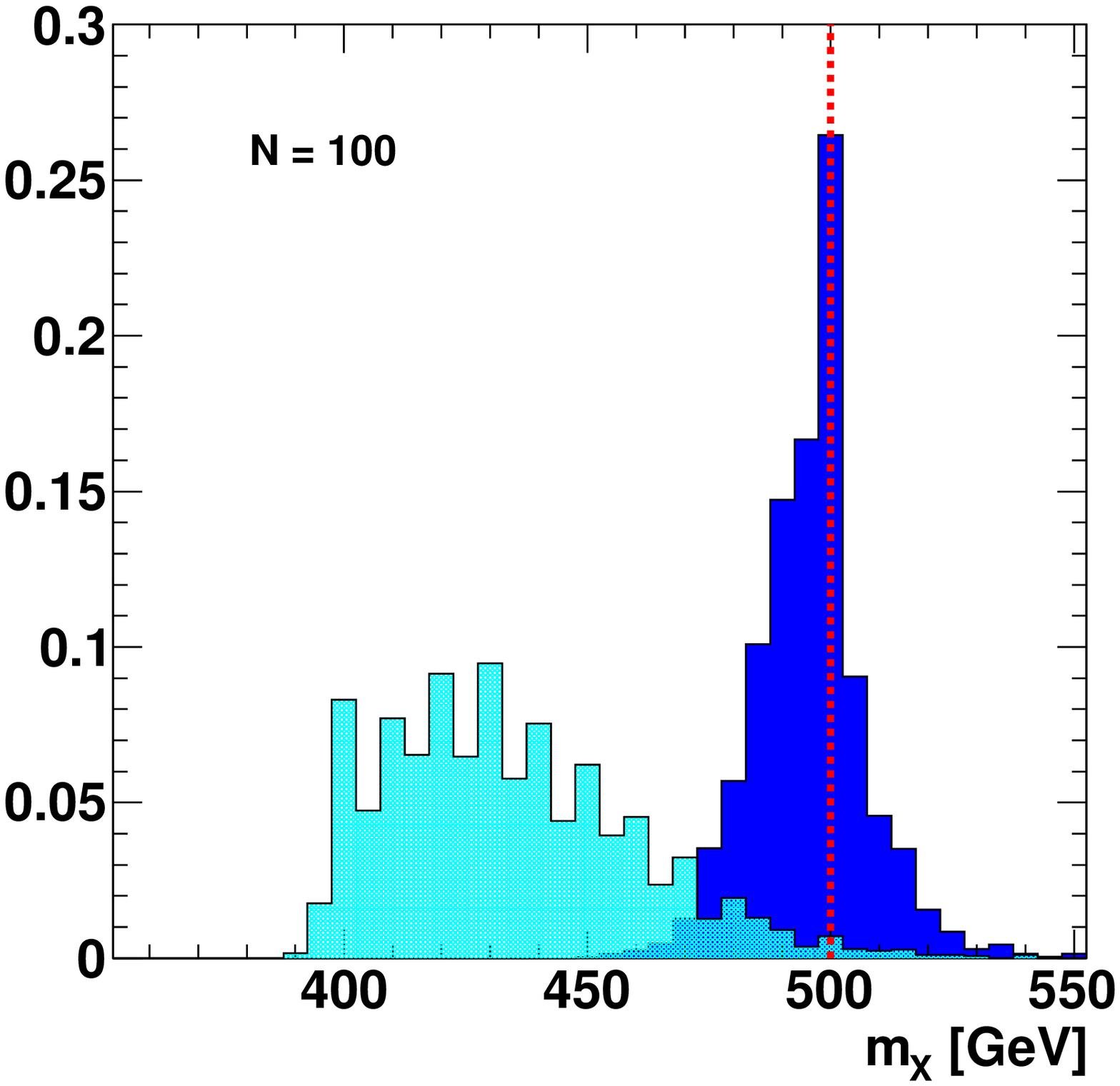}
    \includegraphics[scale=0.25]{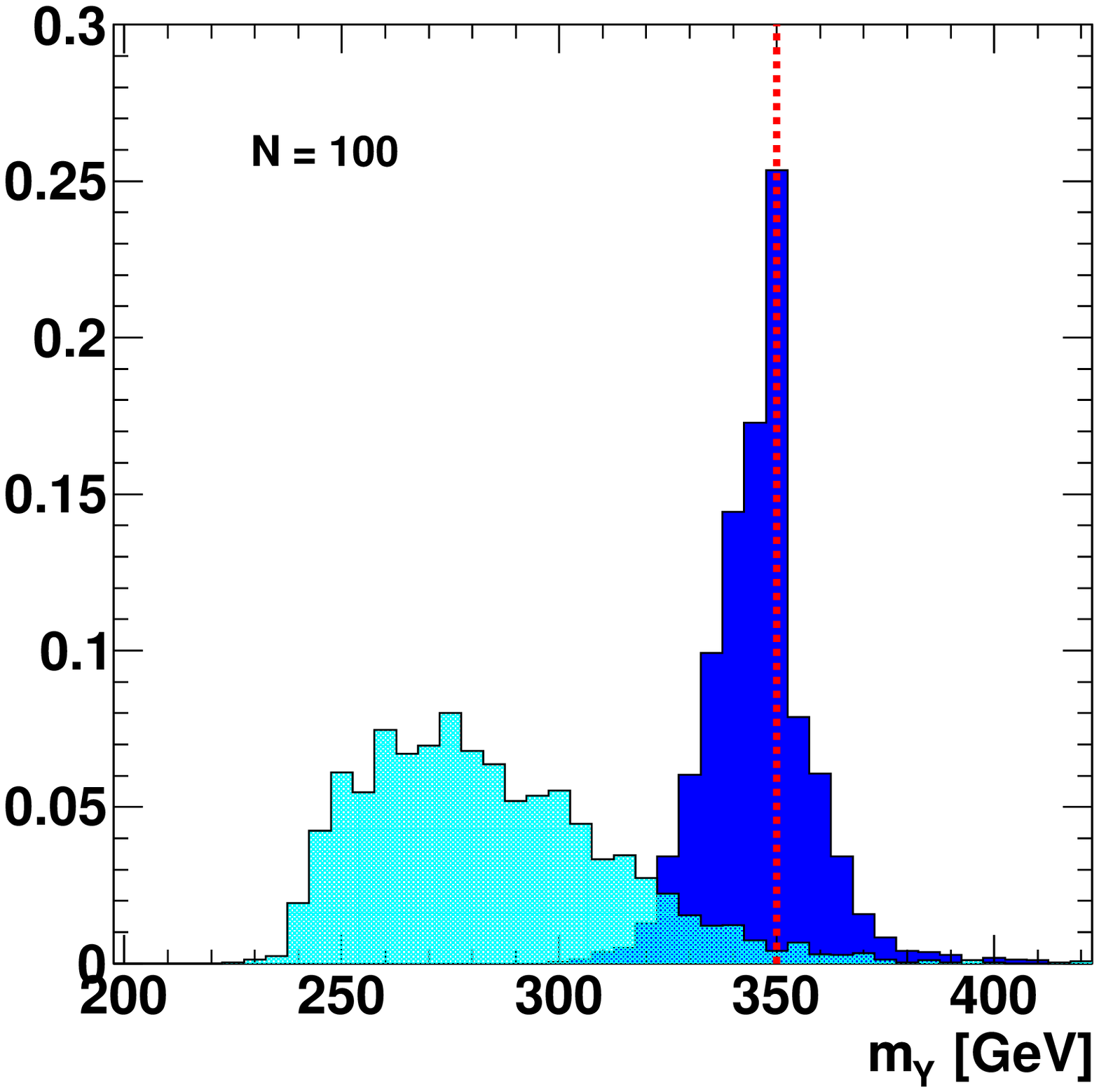}
    \includegraphics[scale=0.25]{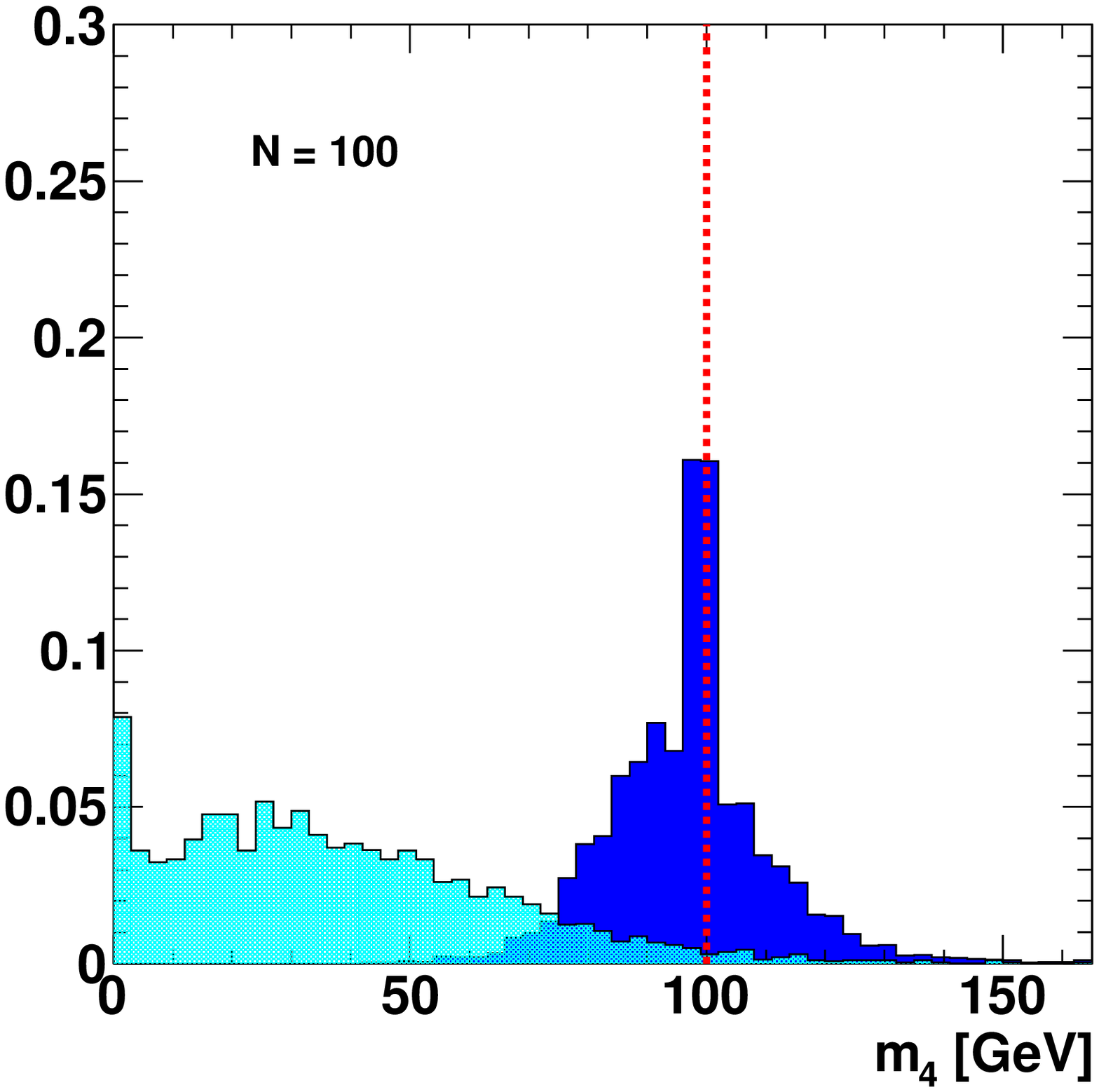}
    }
    \\
    \subfloat[]{
    \includegraphics[scale=0.25]{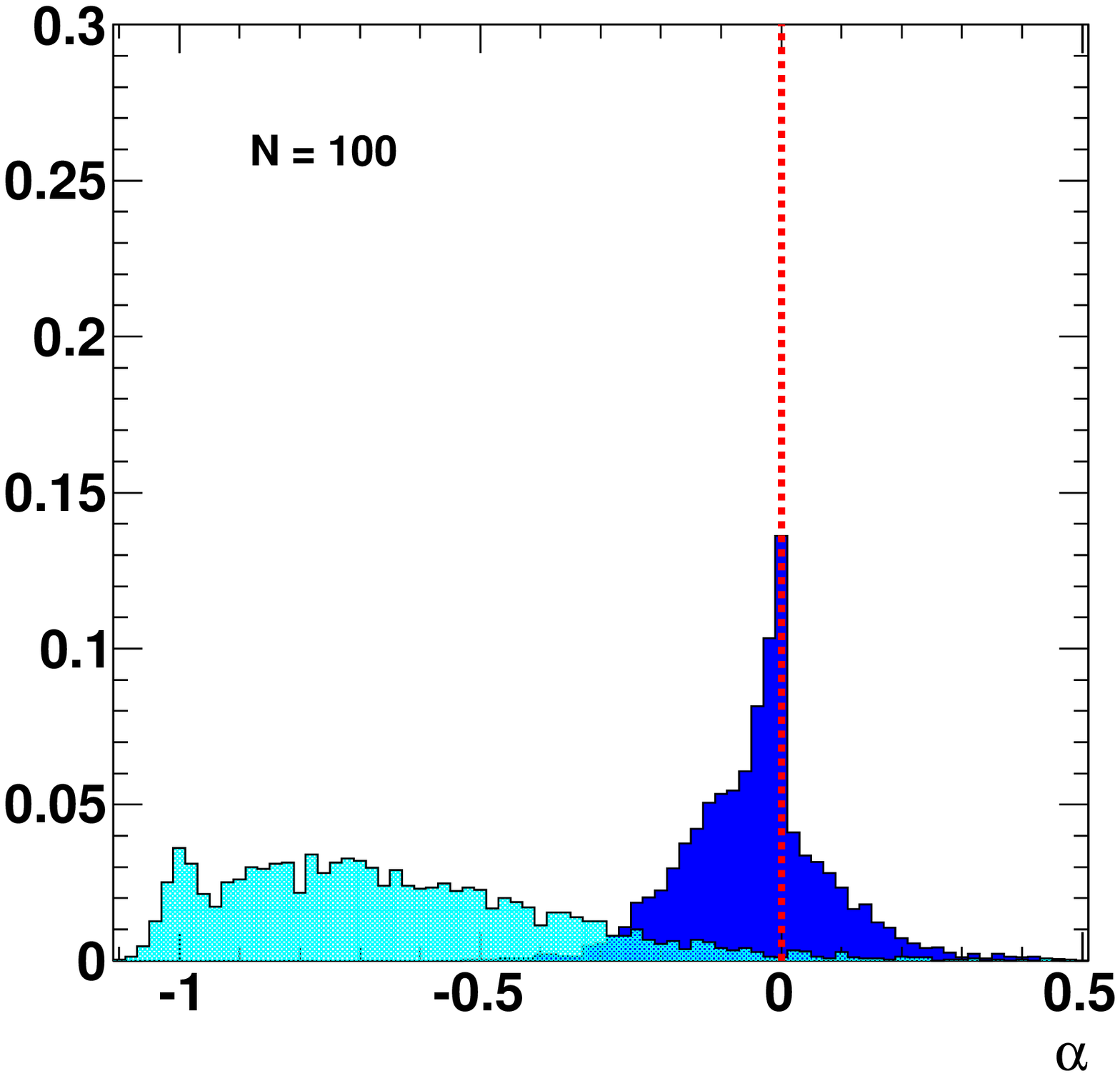}
    \includegraphics[scale=0.25]{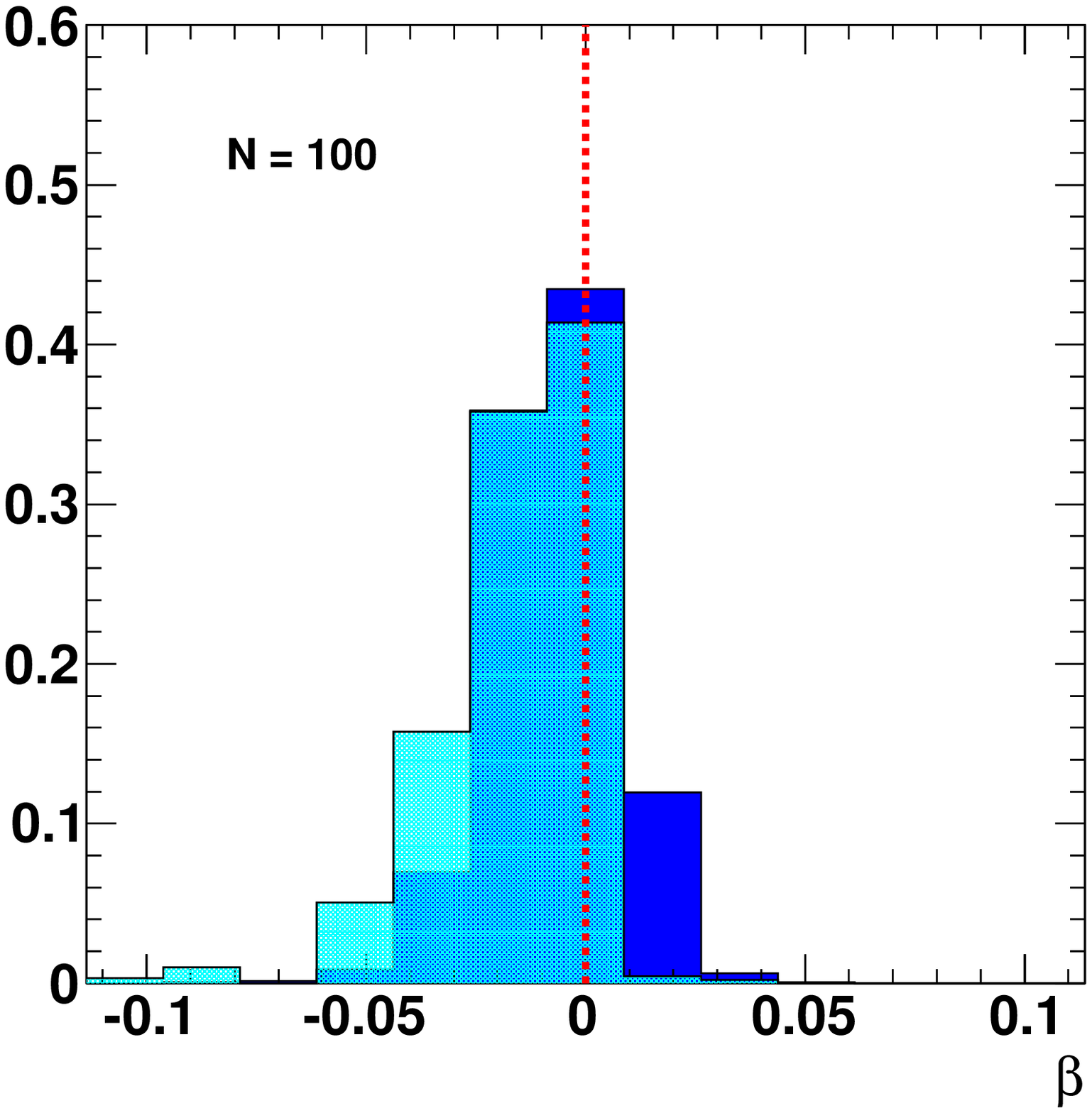}
    \includegraphics[scale=0.25]{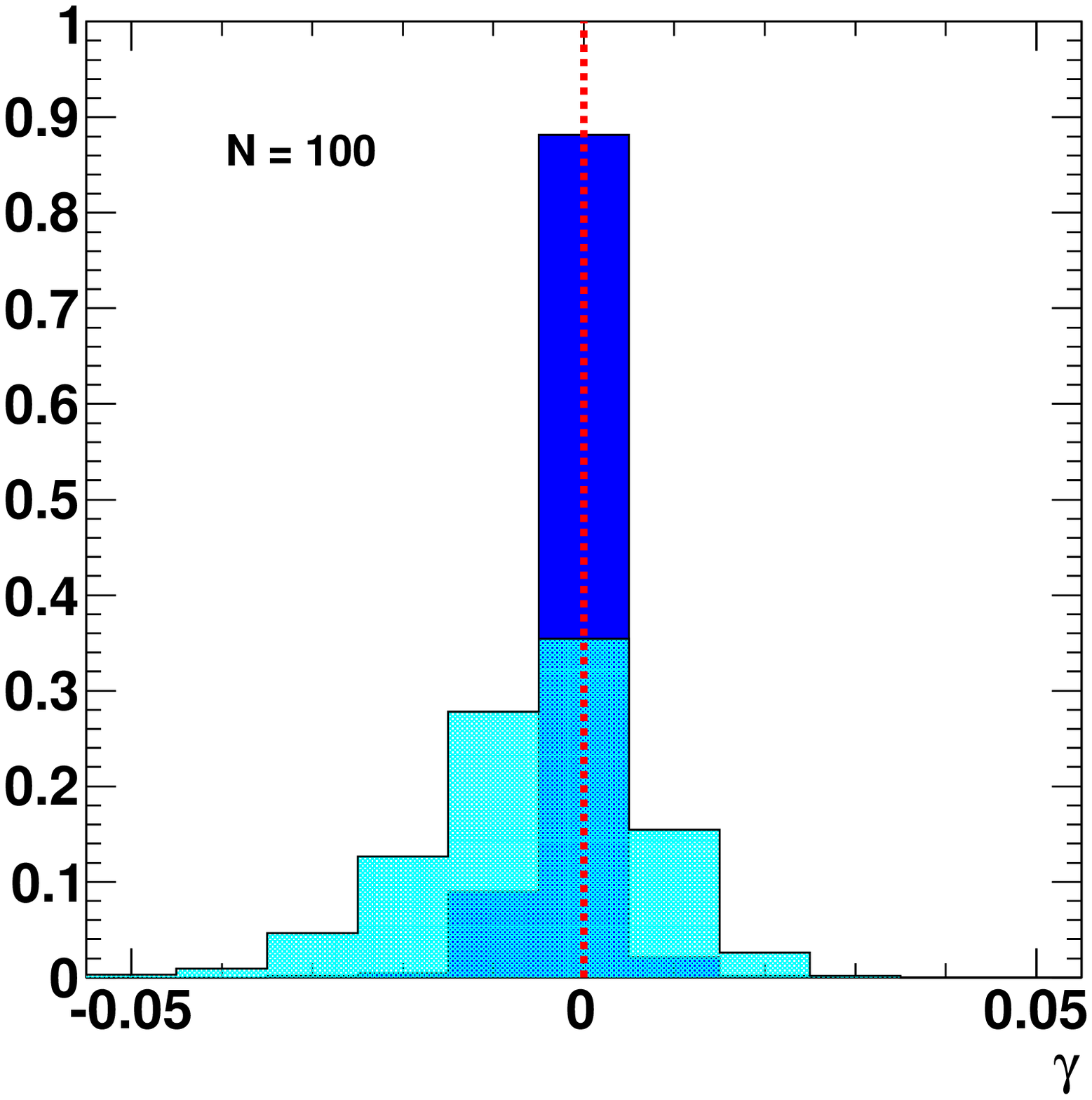}
    }
    \caption{
    The distribution of winning mass hypotheses across pseudoexperiments with 100 events each
    using the phase space analysis (blue) and endpoint analysis
    (cyan) for the
    ``2+3'' topology, both in terms of the physical masses and also parameterized by $(\alpha,\beta,\gamma)$ as explained in the text. The true values in the benchmark spectrum used to
    generate the data are shown by dashed red lines.
    The sawtooth-like shape in the $m_X$ histogram is an artifact of the finite grid size.
    }
    \label{fig:23_100}
  \end{center}
\end{figure}

\begin{figure}[tp]
  \begin{center}
    \subfloat[]{
    \includegraphics[scale=0.25]{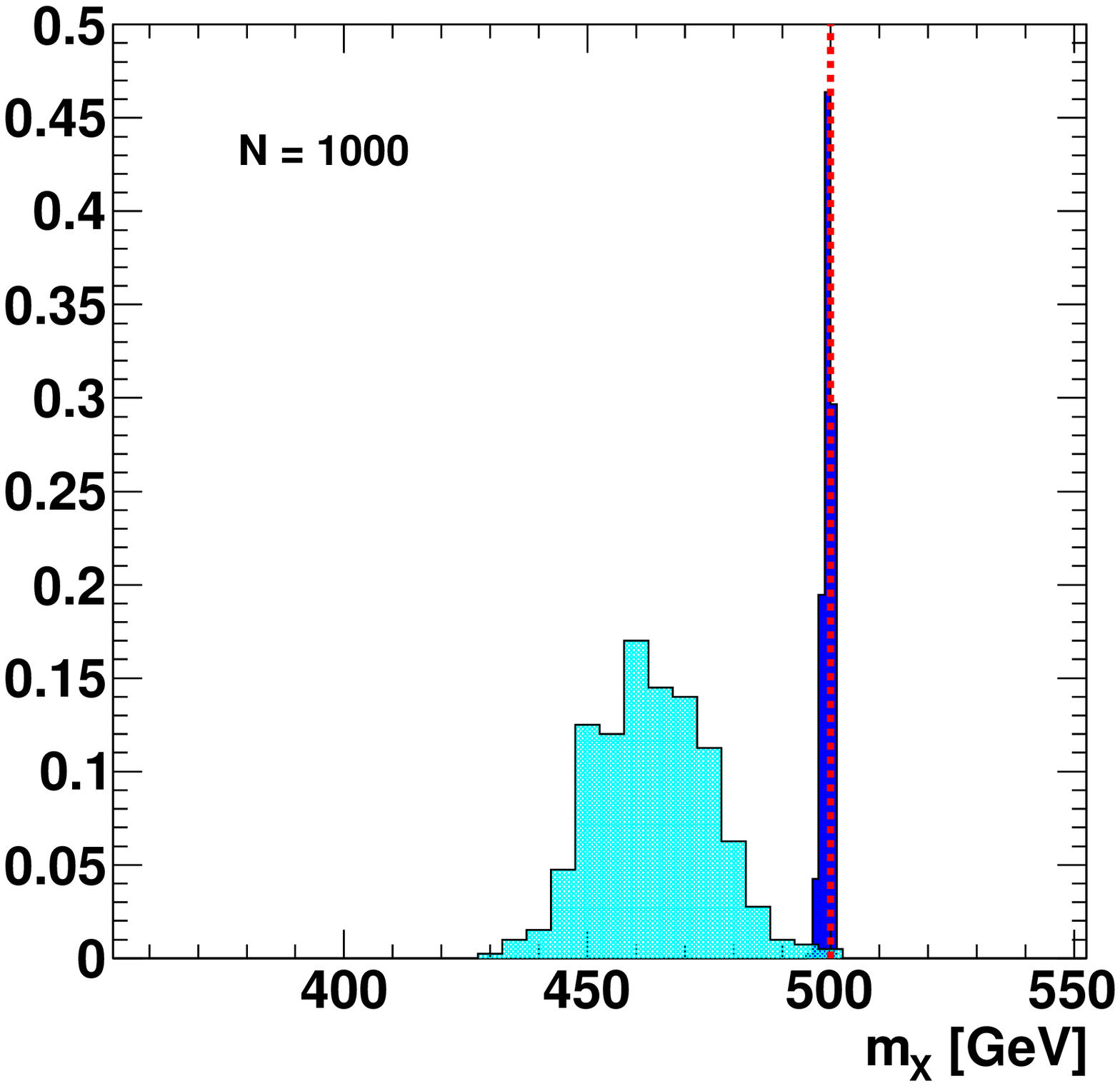}
    \includegraphics[scale=0.25]{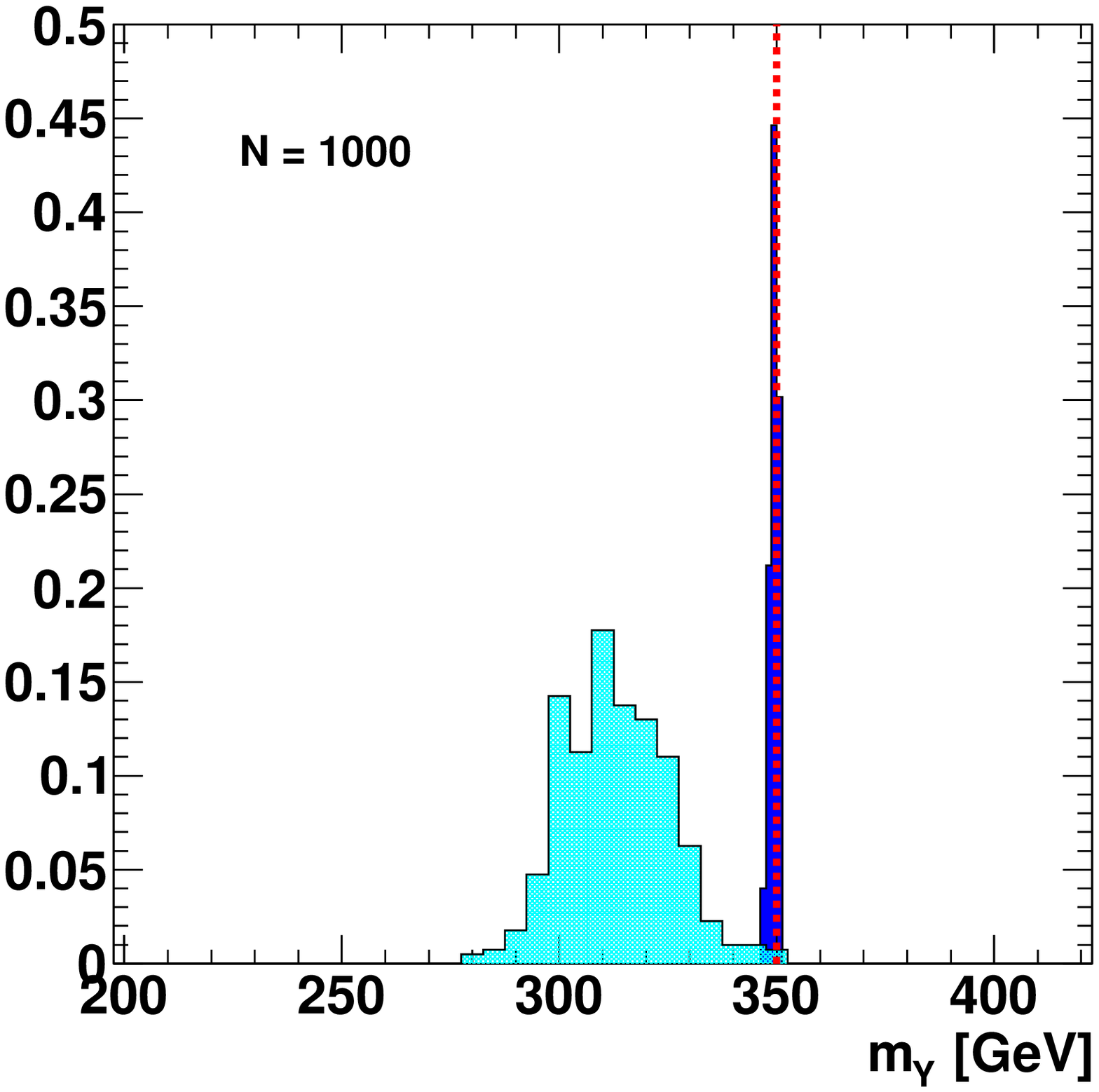}
    \includegraphics[scale=0.25]{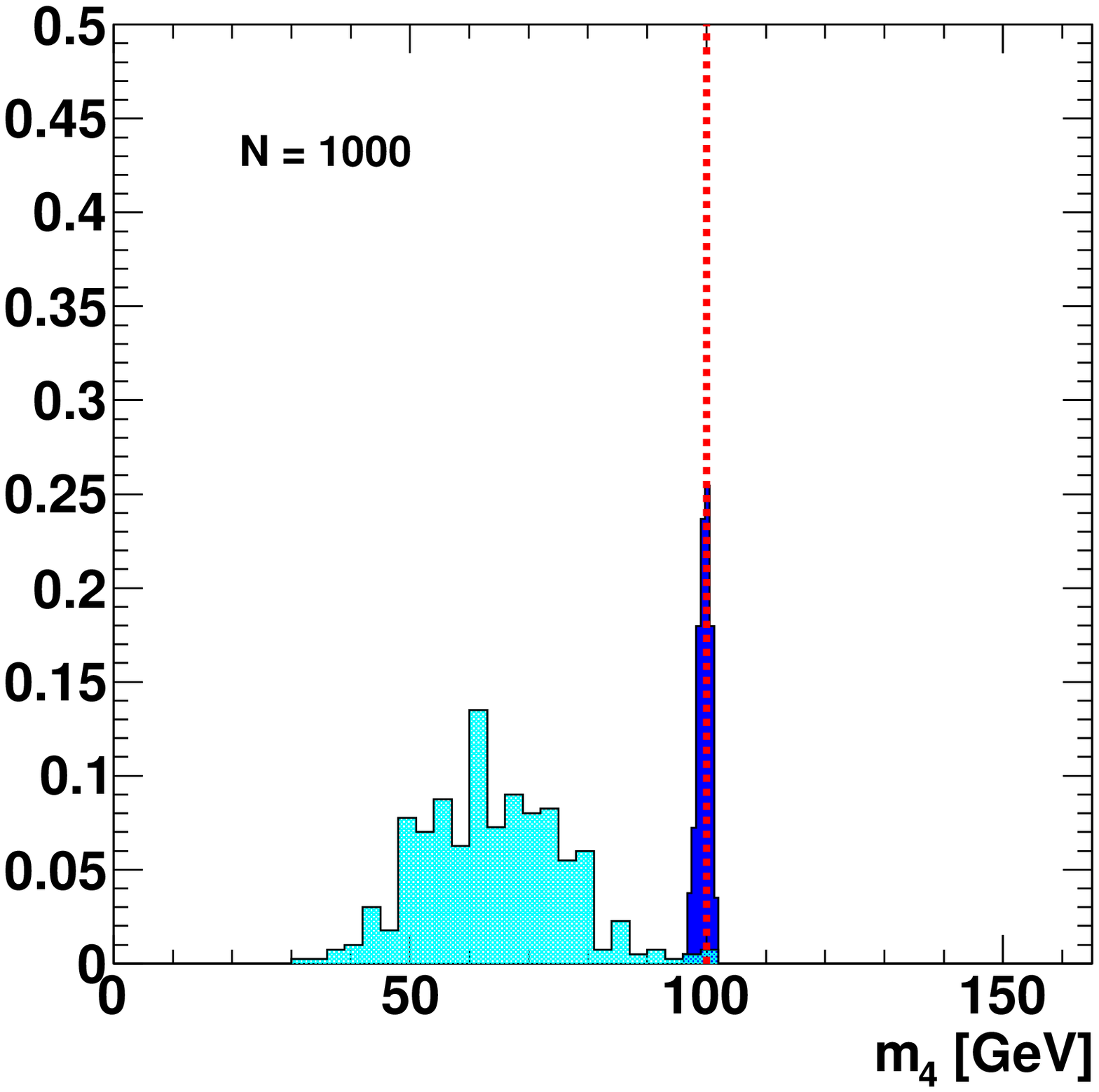}
    }
    \\
    \subfloat[]{
    \includegraphics[scale=0.25]{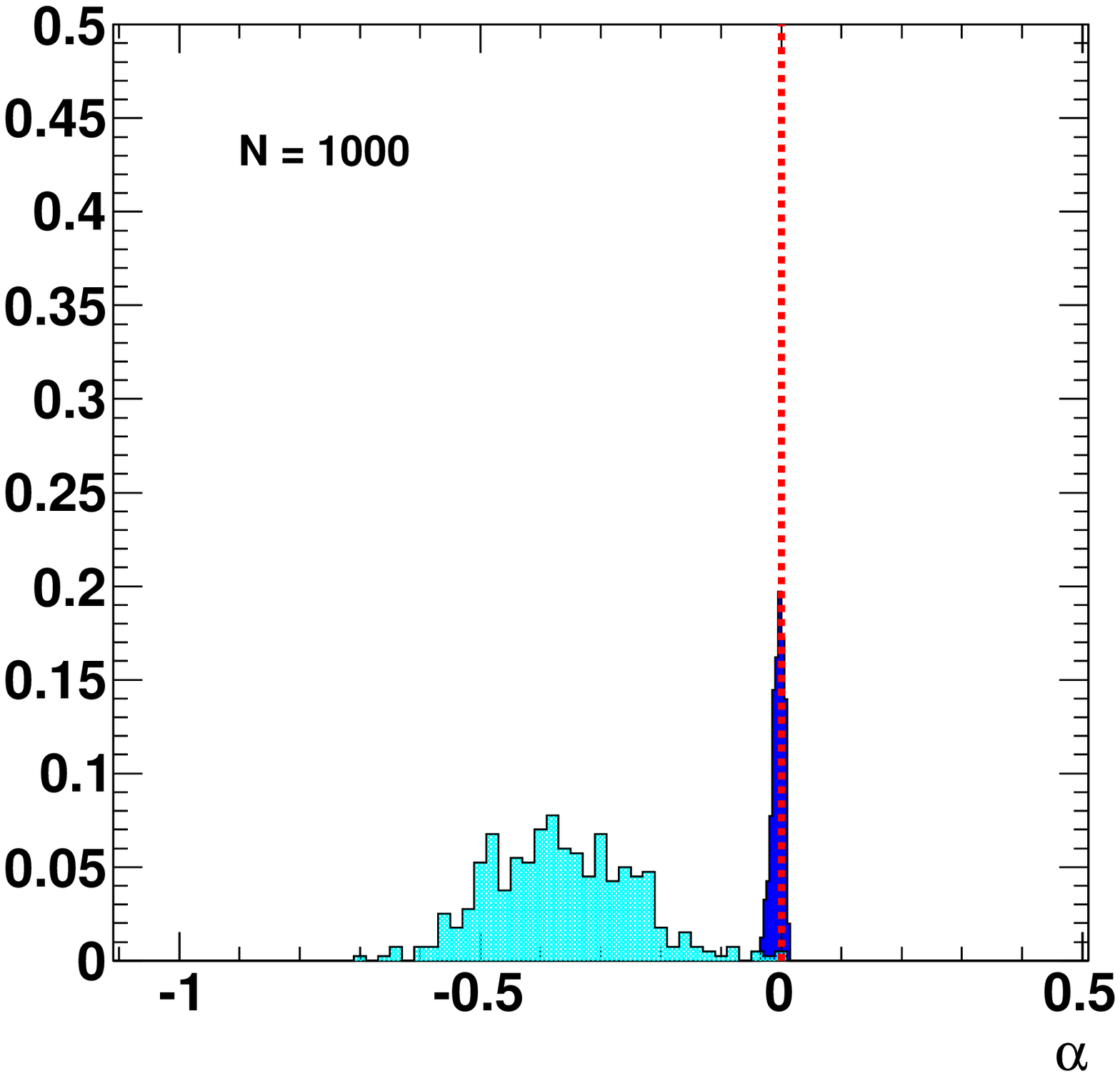}
    \includegraphics[scale=0.25]{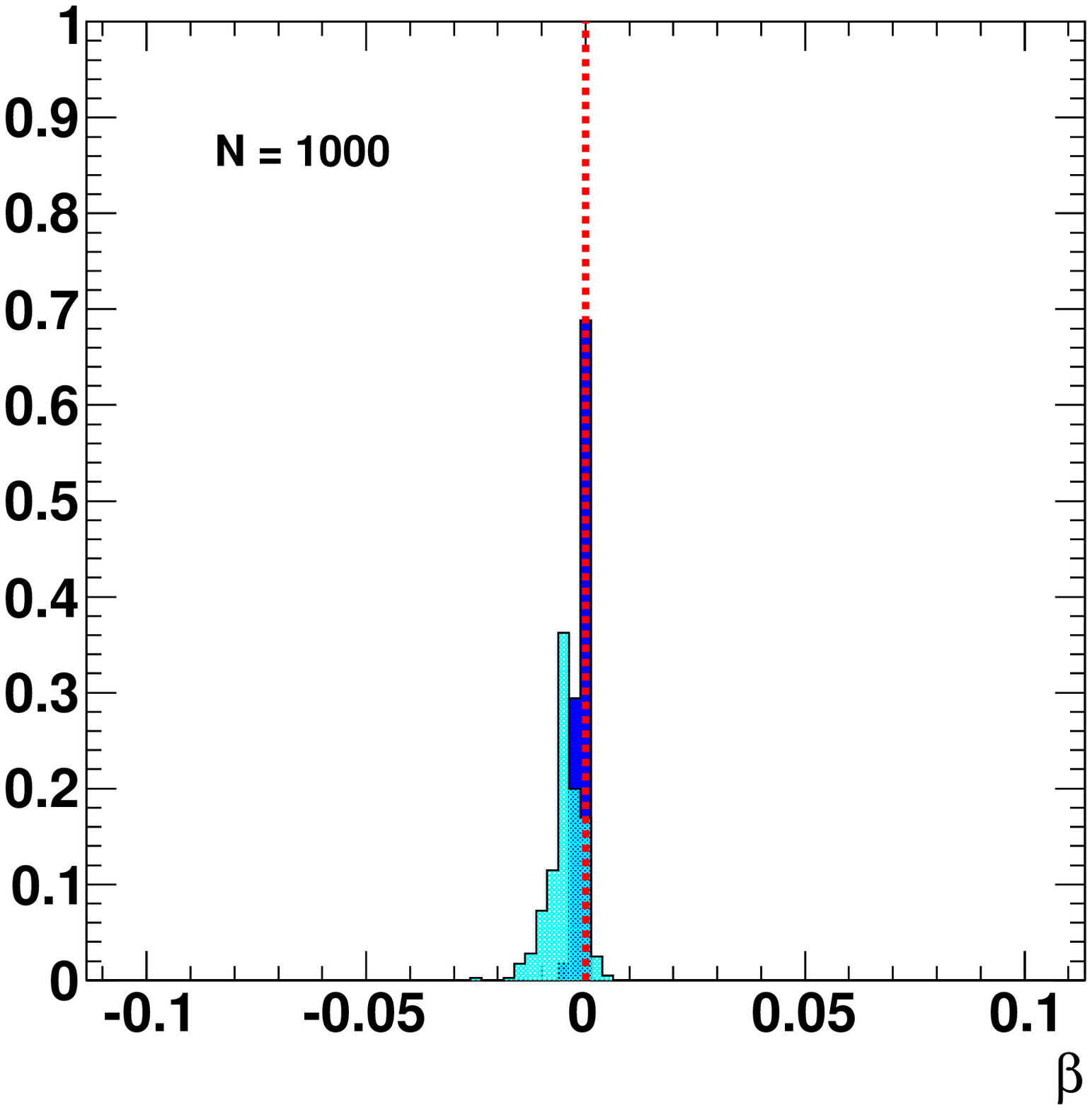}
    \includegraphics[scale=0.25]{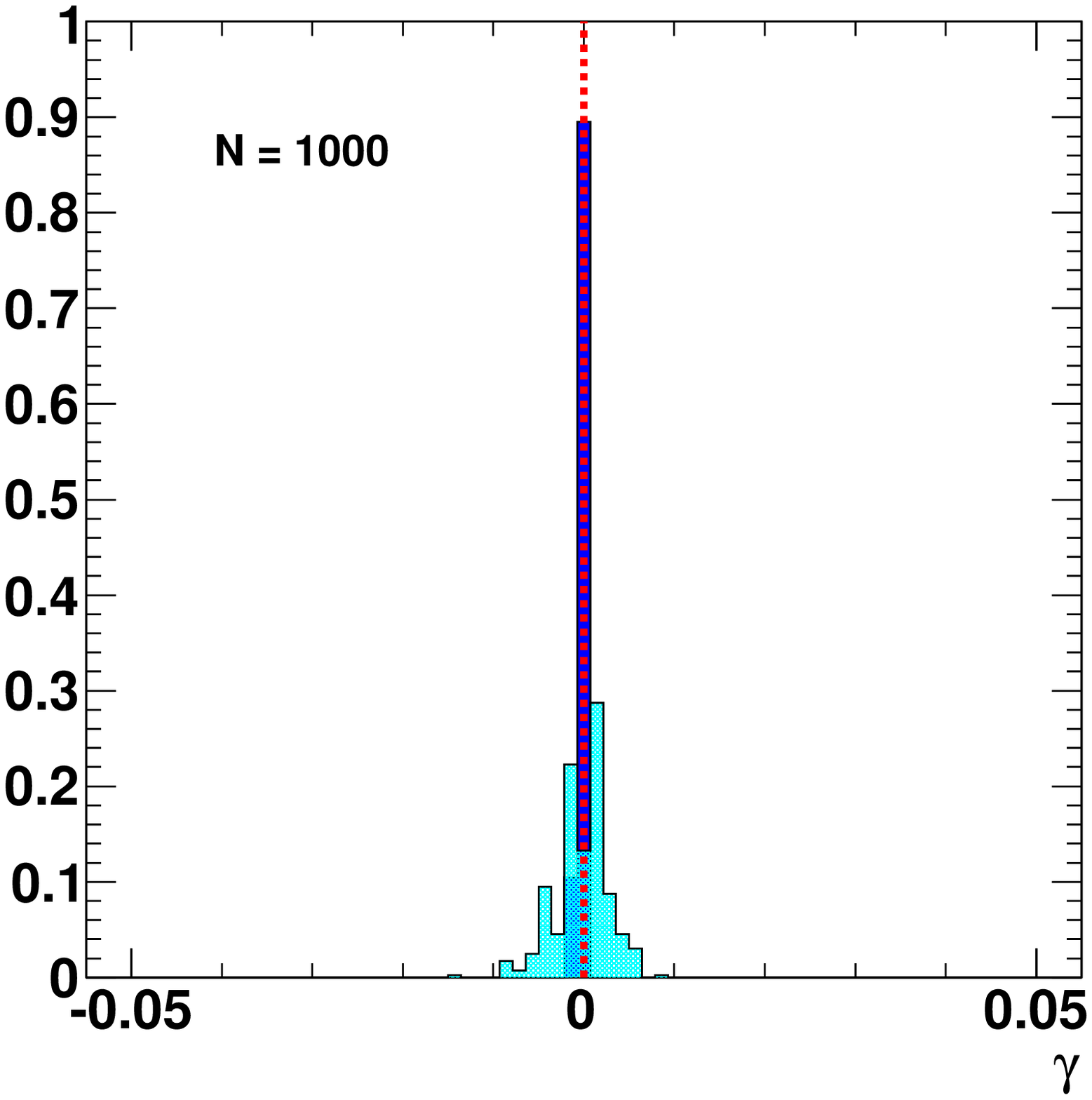}
    }
    \caption{
    The distribution of winning mass hypotheses across pseudoexperiments with 1000 events each
    using the phase space analysis (blue) and endpoint analysis
    (cyan) for the
    ``2+3'' topology, both in terms of the physical masses and also parameterized by $(\alpha,\beta,\gamma)$ as explained in the text. The true values in the benchmark spectrum used to
    generate the data are shown by dashed red lines.
    }
    \label{fig:23_1000}
  \end{center}
\end{figure}

As in the 2+2+2 topology, we run the  phase space and endpoint
analyses on a large number of pseudoexperiments of 100 events each.
Furthermore, in order to study the robustness of our conclusions, we
choose to run a parallel study, this time with 1000 events in each
pseudoexperiment. In figures~\ref{fig:23_100} and \ref{fig:23_1000} we
plot the distribution of winners in these two runs in both the
$(\alpha,\beta,\gamma)$ and $\{\tilde{m}_X,\tilde{m}_Y,\tilde{m}_4\}$
parameterizations. The mean and standard deviations of the mass
parameters in both runs are reported in table~\ref{tab:23}.

Once again, it is very easy to observe stark difference in sensitivity
between the two analysis methods. As in the 2+2+2 topology, the
phase space analysis can resolve the
flat direction. Note that while the spread along the flat direction in
the endpoint analysis does not appear as broad in this case as in
the 2+2+2 topology, this is partly due to the fact that the analysis
is biased towards lower values of $\tilde{m}_{\sigma}$ and thus the
flat direction is cut off by the physical requirement that
$\tilde{m}_{4}\ge0$. This bias can be understood quantitatively, as
discussed in detail in appendix \ref{sec:edge-end-point}.

\begin{table}[tp]
  \begin{tabular}{|c|c|c|c|c|}
    \hline\hline
    \multirow{2}{*}{Mass (GeV) }&

    \multicolumn{2}{|c}{$N_{events}=100$} &
    \multicolumn{2}{|c|}{$N_{events}=1000$} \\
    \cline{2-5}
     & Phase space & Endpoints
     & Phase space & Endpoints \\
    \hline
    $m_{X}$  & $495.84 \pm 11.95$ & $434.32 \pm 25.93$ &$499.40 \pm
    0.96$ &$463.32 \pm 11.66$\\
    $m_{Y}$  & $345.69 \pm 12.13$ & $284.11 \pm 28.48$ &$349.39 \pm
    0.97$ &$312.94 \pm 12.08$\\
    $m_{4}$  & $ 96.86 \pm 13.97$ & $ 37.61 \pm 27.45$ &$ 99.56 \pm
    1.08$ &$ 63.83 \pm 11.91$\\
    \hline
    $\alpha$ & $-0.039 \pm 0.127$& $ -0.647 \pm 0.272$
    & $(-5.49 \pm 9.97)\times10^{-3}$ & $-0.37 \pm 0.12$\\
    $\beta $ & $-0.006 \pm 0.013$& $ -0.017 \pm 0.020$
    & $(0.89 \pm 1.05)\times10^{-3}$
    & $(-4.4 \pm 3.9) \times 10^{-3}$\\
    $\gamma$ & $-0.001 \pm 0.005$& $ -0.005 \pm 0.012$
    & $ (0.23\pm 0.38)\times10^{-3}$
    & $ (-0.2 \pm 3.0) \times 10^{-3}$\\
    \hline\hline
  \end{tabular}
  \caption{
  Mean and standard deviation of the distribution of winning mass
  hypotheses across all pseudoexperiments using the phase space and
  endpoint analyses for the 2+3 topology, for
  different sample sizes.
  }
  \label{tab:23}
\end{table}

Since we have higher statistics in the $N_{events}=1000$ run,
we refine our scan range for this case.
For the phase space analysis, the scan parameters were
\begin{align}
\alpha &\in [-3.5\times10^{-2},1.0\times10^{-2}] \qquad
\Delta\alpha = 5\times10^{-4}, \nonumber \\
\beta &\in [-4.0\times10^{-3},1.0\times10^{-3}] \qquad
\Delta \beta = 2\times10^{-4},\nonumber  \\
\gamma &\in [-1.5\times10^{-3},0.5\times10^{-3}]  \qquad
\Delta \gamma = 10^{-4}.
\end{align}

For the endpoint analysis, the corresponding numbers chosen were,
\begin{align}
\alpha &\in [-2.0,2.0] &
\Delta \alpha &= 0.02, \nonumber \\
\beta &\in [-4.0\times10^{-2},4.0\times10^{-2}]
& \Delta \beta &= 2\times10^{-3},\nonumber \\
\gamma &\in [-2.0\times10^{-2},2.0\times10^{-2}] &
\Delta \gamma &= 10^{-3}.
\end{align}

Indeed, the endpoint analysis
in this case gives a reasonable estimate of the masses, with a
precision of less than $10\%$, and a smaller bias (within $\sim 3\sigma$ of
the true value). In contrast, however, the endpoint analysis with
$N_{events} = 100$ gives very inaccurate results, whereas the
phase space analysis still performs well. This makes it clear that
while the phase space analysis is still preferred at high statistics
due to reduced errors, it becomes indispensable at low statistics.

\section{Outlook and Conclusions} \label{sec:conclusions}
Given that the LHC has not already observed signs of new physics, there is a
significant probability that even if we find new physics, we will have only a
small number of signal events available to study its properties. We have
demonstrated that in a low statistics environment, mass determination in
cascade decays can be significantly improved by determining the boundary of the
kinematically accessible phase space in its full dimensionality. This yields
better results than traditional methods based on the presence of edges and
endpoint in one-dimensional kinematic distributions because:
\begin{enumerate}
{\item There are correlations in the distribution of events in the full phase space that are lost when the phase space is projected onto any one-dimensional observable.}
{\item For cascade decays with more than three final state particles, the phase space density is enhanced near the edge of the kinematically accessible region, allowing one to map the boundary with better resolution and utilize the aforementioned correlations.}
\end{enumerate}

By focusing on an event topology with one particle decaying to three visible
and one invisible final state particle through a two-stage decay, we were able
to demonstrate this improvement in mass determination quantitatively. Not only
were the the best fit values for the unknown masses found to be closer to the
true values used in the generation of the data, but the uncertainties in the
measurement were also shown to be smaller when the analysis was performed on
the full phase space in a likelihood-based analysis. Finally, while mass
determination methods based on kinematic edges and endpoints exhibit a
weakness in that they are sensitive to differences of masses in the
spectrum but insensitive to the overall scale, the phase space analysis
allows one to measure all unknown masses in the decay chain accurately.

While we demonstrated the applicability of our results on a reasonably simple event topology, our conclusions have wider applicability. The polynomial method may fail to apply even for a longer decay chains, if the cascade proceeds through three-body decays, in which case the system remains underconstrained. The phase space methods on the other hand become more powerful in this case since the shape of the phase space boundary in higher dimensions carries all the necessary information about the masses of the unknown particles occurring in the cascade. It should be noted however that the functional form of the phase space density is complicated for more than four final state particles and we leave the study of a more general class of decay topologies to future work.

In conclusion, we wish to briefly address future directions of study and also
dwell on how the simplifying assumptions made in this study can be relaxed such
that the validity of our conclusions can be tested in a more realistic collider
analysis.

\begin{itemize}

\item
The phase space analysis can be extended to cases where both sides of the event contain a cascade. The main added complication in this case is the inclusion of the additional constraint due to the measurement of the missing transverse energy (MET) into the analysis. This can be done in terms of
Lorentz-invariant variables by introducing ``reference'' momenta (for example the four-momenta of the beams) which can be used in order to construct projection operators onto the transverse plane.

For cascades on either side of the event which have fewer than four
final state particles (one of which is invisible), there is at most
one Lorentz invariant observable for that side of the event, thus it
is not possible to go beyond an edge or endpoint analysis on either
side alone. It is in these cases where it may be of interest to
consider the full phase space distribution of the entire event, rather
than each side of the event separately, which necessitates the
inclusion of the MET constraint into the likelihood method. In the
extreme case of only one visible final state on either side, there is
only a single Lorentz invariant observable for the entire event.  This
is consistent with the conclusion of ref.~\cite{Cheng:2008hk} that a
single variable, $m_{T2}$, captures all available information about
the kinematics for this decay topology.

\item
While we have focused solely on the properties of the phase space density
in this work, given a particular model, it is straightforward to incorporate
the full matrix element for specific processes as well into the likelihood analysis
as well.  Such an analysis will not only differentiate between mass hypotheses,
but it can also yield information about the spins and interactions of the
particles involved.

\item
In our study, we have assumed an ideal detector with perfect energy resolution. As a result, even a single event that were incompatible with a given mass hypothesis could rule it out. In a study with a more realistic detector model, the effects of finite energy resolution will need to be taken into account. This can be accomplished by convoluting the likelihood with the response of the detector. The likelihood will then have a tail beyond the boundary of phase space. This will almost certainly reduce the precision of the mass measurement, however it should be kept in mind that a similar degradation will occur in the edge and endpoint based measurements as well.

\item
We have ignored the issue of combinatoric ambiguity. This can be particularly unrealistic when there are a large number of identical final states. The definition of the likelihood can be extended to sum over all possible ways that the observed particles can be assigned to the final state particles in the decay topology.

{\item
We have assumed that the correct decay topology is used in the analysis. One can imagine relaxing this assumption without greatly modifying the analysis, by comparing the likelihoods based on a range of applicable decay topologies. This way, a likelihood analysis based on the full dimensionality of phase space can be used effectively not only to measure the masses of unknown particles but also to determine the correct underlying decay topology among a number of alternatives.}

{\item
We have ignored the presence of SM backgrounds in our study. In order to remove this assumption, one needs to model the background distribution in the multidimensional phase space and assign a likelihood to each event as having arisen from signal or from background based upon this model. Similar statistical techniques are used in likelihood-based analyses in new physics searches by ATLAS and CMS, though in a different context. It may in general be difficult to obtain an accurate model of the full background distribution, which will reduce the precision of the mass measurement. While a degradation will also occur as background is included in edge and endpoint based analyses, the degree to which the two analyses will be impacted is difficult to predict. On one hand, a multidimensional background model is more arbitrary and may therefore lead to greater loss of precision in the result of the mass measurement. On the other hand, due to the enhancement of the signal phase space distribution near the boundary of phase space, which is absent for the background, the details of the background model away from the boundary may not be as important as it would at first appear. One way to visualize this is to imagine an extreme limit where the signal events all lie on the boundary, while the background will vary smoothly in the direction orthogonal to the boundary, but the precise way in which the distribution of the background is modeled along this direction would not be important.

}

\end{itemize}

A more detailed study is necessary in order to more exactly quantify the impact of removing these simplifying assumptions on the precision of the mass measurements based on the likelihood based method vs. the edge and endpoint based methods. This will be further explored in future work.

\begin{acknowledgments}
We thank Kuver Sinha for collaboration in the early phases
of this work. PA would like to thank Ciaran Williams for helpful
discussion.  The research of CK and JHY was supported in part by NSF
grant PHY-0969020.  Fermilab is operated by Fermi Research Alliance,
LLC under Contract No. DE-AC02-07CH11359 with the United States
Department of Energy. CK would like to thank the Aspen Center for Physics where part of this work was completed (supported by the National Science Foundation under Grant No. PHYS-1066293). CK would also like to thank the Kavli Institute for Theoretical Physics where part of this work was completed (supported by the National Science Foundation under Grant No. NSF PHY11-25915).

\end{acknowledgments}

\appendix
\section{Formulae for edges and end-points}\label{sec:edges-endpoint-form}
In this appendix we present the list of analytical formulae for the position of edges and
endpoints in one-dimensional kinematic distributions that we use in our analysis. These have been well studied in the literature
\cite{Hinchliffe:1996iu,Allanach:2000kt,
Gjelsten:2004ki,Gjelsten:2005aw,Miller:2005zp,Burns:2009zi,Matchev:2009iw}.
Throughout this section, we work in the approximation $m_1 = m_2 = m_3 \simeq
0$. We have verified numerically that this approximation is valid for our
purposes.

\subsection{2+2+2 Topology}

\begin{align}
  (m_{123}^2)_{max}
  &=
  \begin{cases}
    \frac{(m_X^2-m_Y^2)(m_Y^2-m_4^2)}{m_Y^2 }
    & \frac{m_X}{m_Y} > \frac{m_Y}{m_Z} \frac{m_Z}{m_4} \\
    \frac{(m_X^2 m_Z^2 - m_Y^2 m_4^2)
    (m_Y^2-m_Z^2)}
    {m_Y^2 m_Z^2}
    & \frac{m_Y}{m_Z} > \frac{m_Z}{m_4} \frac{m_X}{m_Y} \\
    \frac{(m_X^2-m_Z^2)(m_Z^2-m_4^2)}{m_Z^2 }
    & \frac{m_Z}{m_4} > \frac{m_X}{m_Y} \frac{m_Y}{m_Z} \\
    (m_X - m_4)^2
    & \mathrm{otherwise}
  \end{cases}
\end{align}
\begin{align}
  (m_{23}^2)_{max}
  &=
  (m_Y^2-m_Z^2)(m_Z^2-m_4^2)/m_Z^2,
\end{align}
\begin{align}
  (m_{12}^2)_{max}
  &=
  (m_X^2-m_Y^2)(m_Y^2-m_Z^2)/m_Y^2,
\end{align}
\begin{align}
  (m_{13}^2)_{max}
  &=
  (m_X^2-m_Y^2)(m_Z^2-m_4^2)/m_Z^2 \,.
\end{align}


In cases where some of the final state particles are not distinguishable, it can be more convenient to use endpoints in the distribution of the higher and lower invariant mass of certain combinations of final states. For example, if the visible final states contain one lepton and two jets, the two possible lepton-jet pair invariant masses can be labeled event-by-event as $m^{2}_{lj,high}$ and $m^{2}_{lj,low}$, rather than $m^{2}_{lj_{1}}$ and $m^{2}_{lj_{2}}$. Since we assume that we can identify the final states, this issue does not arise in our study.

\subsection{2+3 Topology}

Since for the 2+3 Topology, there is no intrinsic distinction that sets particles 2 and 3 apart, the kinematic distributions respect the discrete symmetry $2\leftrightarrow3$.

\begin{align}
  (m_{123}^2)_{max}
  &=
  \begin{cases}
    \frac{(m_X^2-m_Y^2)(m_Y^2-m_Z^2)}{m_Y^2 }
    & \frac{m_X}{m_Y} > \frac{m_Y}{m_4}\\
    (m_X - m_4)^2
    & \mathrm{otherwise}
  \end{cases}
\end{align}
\begin{align}
  (m_{23}^2)_{max}
  &=
  (m_Y-m_4)^2,
\end{align}
\begin{align}
  (m_{12}^2)_{max}
&=
  (m_{13}^2)_{max} =
  (m_X^2-m_Y^2)(m_Y^2-m_4^2)/m_Y^2 \,.
\end{align}

\section{$m_{34}^{2}$ Integral for the 2+3 Topology}
\label{sec:integrating-m34}
In this appendix we provide additional details about how to analytically take the $m_{34}^{2}$ integration necessary for the likelihood analysis in the 2+3 topology. The integral is
performed most easily by switching from the $m_{ij}^{2}$ variables to the set $\{m_{12},m_{23},m_{123},m_{34}\}$, which are linearly related to each other. In this basis,
$\Delta_4$ is quadratic in $m_{34}^2$, and can be therefore written as
\begin{align}
  \Delta_4 &= \frac{1}{16} ( a m_{34}^4 + b m_{34}^2 + c ),
  \nonumber\\
  &= \frac{1}{16} a( m_{34}^2 - s_{34}^+ )( m_{34}^2 - s_{34}^- ),
\end{align}	
where,
\begin{align}
  s_{34}^\pm = - \frac{b}{2a} \pm \frac{\sqrt{b^2 -4 a c}}{2a}.
\end{align}
Since $\Delta_4 > 0$ defines the physical phase-space region,
$s_{34}^{\pm}$ determine the range of the $m_{34}^{2}$ integration.

The constants $a$ and $b$ appearing above are,
\begin{align}
  a &= \lambda\left(m_1^2, m_{23}^2, m_{123}^2\right), \\
  b &=  2\ \mathrm{Det}
  \left|\begin{array}{ccc}
    2 m_{23}^2 & m_{123}^2 + m_{23}^2 - m_1^2 & m_{23}^2 - m_{Y}^2 + m_{4}^2\\
    m_{123}^2 + m_{23}^2 - m_1^2 & 2 m_{123}^2 & m_{123}^2 - m_{X}^2 + m_{4}^2\\
    m_{23}^2 - m_2^2 +m_3^2 & m_{123}^2 -m_{12}^2 + m_3^2 & m_3^2 +m_4^2
  \end{array}\right|,
\end{align}
\begin{align}
  b^2 -4 a c
  &= 16\ G
  \left(m_{12}^2, m_{23}^2, m_{123}^2, m_2^2, m_1^2, m_3^2\right)
  G\left(m_{123}^2, m_{Y}^2, m_{X}^2, m_{23}^2, m_{1}^2, m_4^2\right),
\end{align}
where the kinematic functions $\lambda$ and $G$ were defined in equation~\ref{eq:lambdaG}.

Using this expression for $\Delta_{4}$ quadratic in $m_{34}^{2}$ it is not too difficult to verify that equation~\ref{eq:23likelihoodB} follows from equation~\ref{eq:23likelihoodA}.

\section{Understanding the edge/end-point results}
\label{sec:edge-end-point}
This appendix is devoted to interpreting the performance of the endpoint analysis, and in particular its insensitivity along the flat direction parametrized by one parameter, say
$\tilde{m}_4$ with the other masses chosen as $\tilde{m}_{\sigma} - \tilde{m}_4 = M_{\sigma} - M_{4}$ where $M_{\sigma}$ correspond to the true mass point used to generate the data.
Understanding this flat direction will also provide insight on the bias in the mass measurement that we observe in the endpoint analysis. In this appendix we will study the flat direction for the specific analysis that we describe in the main body of the paper but for a more general understanding of the flat direction the reader is also encouraged to consult references~\cite{Cho:2007qv,Gripaios:2007is,Cho:2007dh,Barr:2007hy,Barr:2009jv,Burns:2008va,Burns:2009zi}.

\begin{figure}[tp]
  \begin{center}
    \subfloat[]{
    \includegraphics[width=0.45\textwidth]{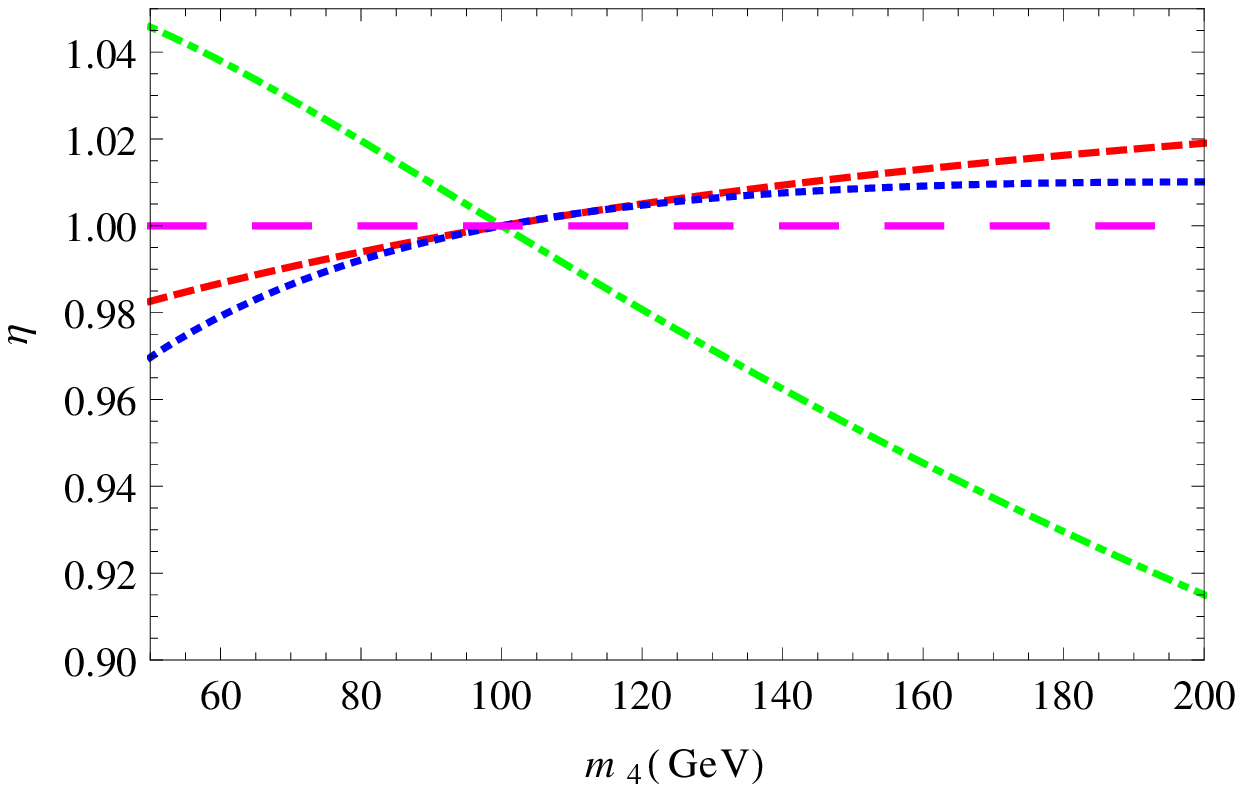}
    }
    \quad
    \subfloat[]{
    \includegraphics[width=0.45\textwidth]{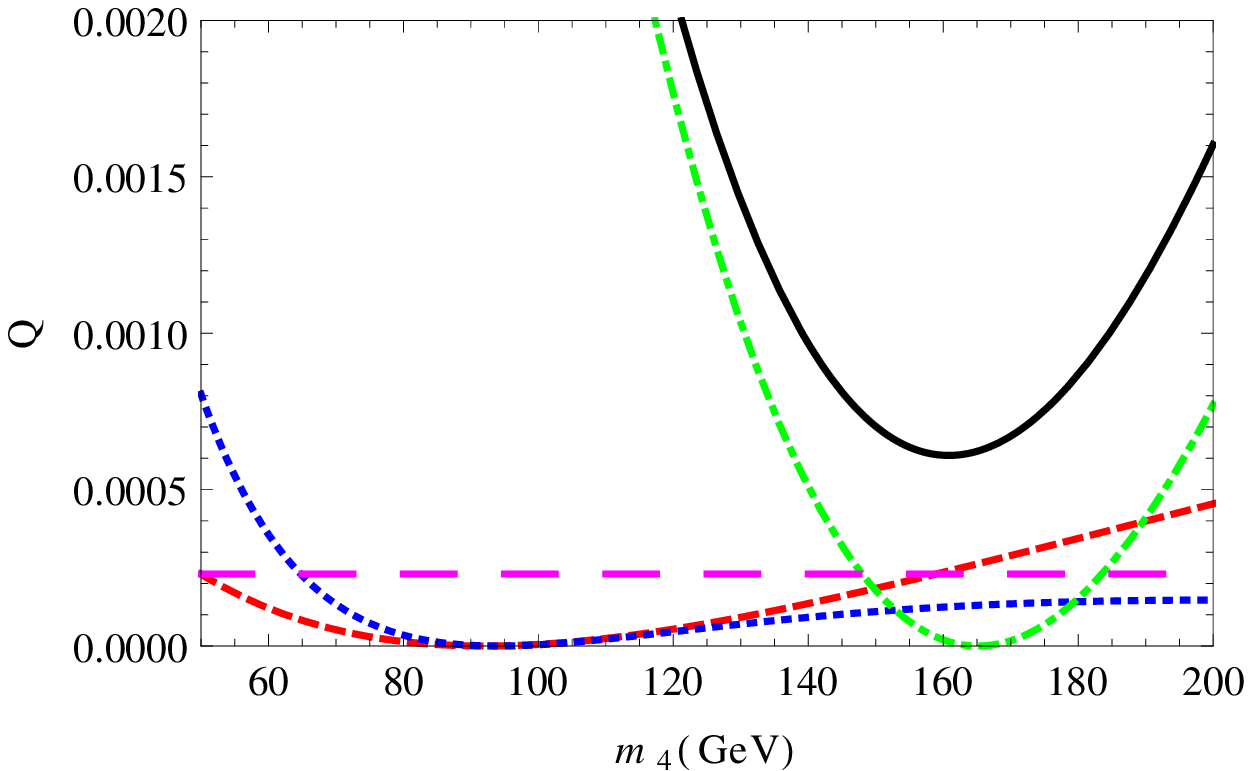}
    }
  \end{center}
  \caption{a) Ratio of predicted values (based on a mass hypothesis along the flat direction) to the true values of the endpoints in $m^2_{12}$ (red, dashed), $m^2_{23}$ (blue, dotted), $m^2_{13}$ (green, dot-dashed) and $m^2_{123}$ (magenta, long-dashed) in the 2+2+2 topology. b) The quality-of-fit variable (defined in equation~\ref{eq:qualityoffit}) for each endpoint along the flat direction. A representative pseudoexperiment is chosen for illustration. The solid black curve corresponds to the total quality-of-fit variable $Q$ for the same pseudoexperiment.  }
  \label{fig:distributions}
\end{figure}

\subsection{2+2+2 Topology}

In order to estimate the dependence of various edge positions along the flat
direction, we plot the ratio of the predicted endpoint position as calculated
with hypothetical masses to the true endpoint position, as a function of
$\tilde{m}_4$. As can be seen from Figure \ref{fig:distributions}, the
endpoints are relatively insensitive to variation of masses along this
direction. In fact, the mass measurement is driven by $(m_{13}^2)_{max}$.

Moreover, in a low statistics sample, kinematic distributions are
often not well populated at the maximal allowed values. This is
especially the case for endpoints (rather than edges), where the
approach to the maximal value is rather shallow. We use a representative
pseudoexperiment to investigate this effect quantitatively. From figure
\ref{fig:distributions}, it is clear that a lower value for $m^2_{13}$ biases
the winning hypothesis towards higher values of $m_4$. This has the effect of
pushing the entire best fit spectrum to higher values.


\begin{figure}[tp]
  \begin{center}
    \includegraphics[width=0.45\textwidth]{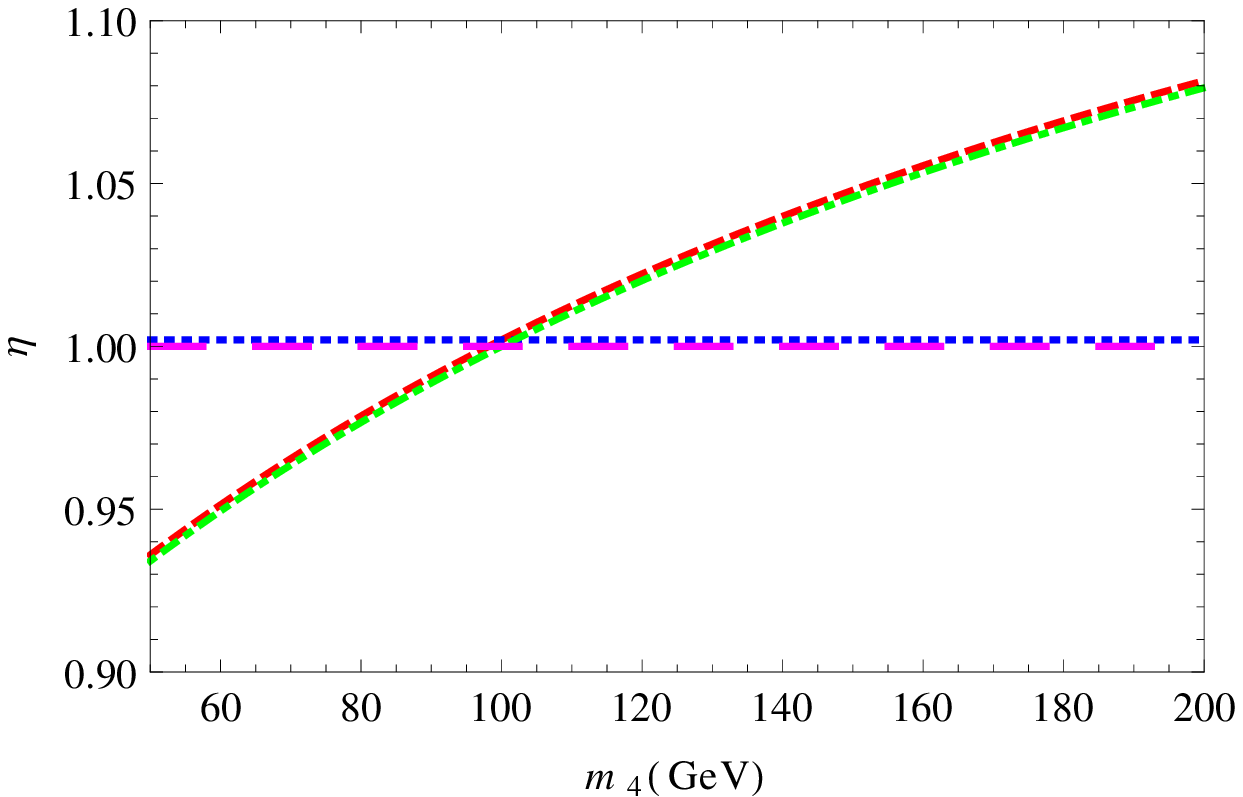}
    \quad
    \includegraphics[width=0.45\textwidth]{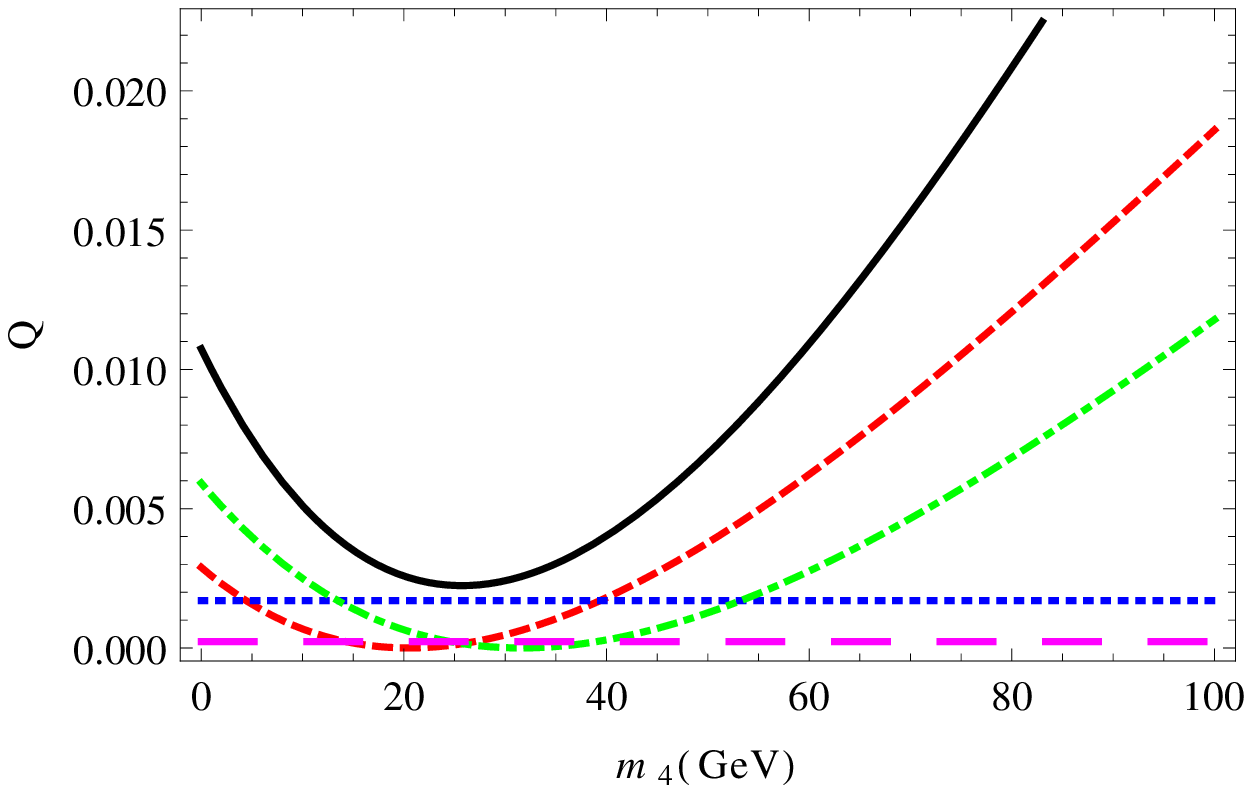}
  \end{center}
  \caption{ a) Ratio of predicted values (based on a mass hypothesis
  along the flat direction) to the true values of the endpoints in
  $m^2_{12}$ (red, dashed), $m^2_{23}$ (blue, dotted), $m^2_{13}$
  (green, dot-dashed) and $m^2_{123}$   (magenta, long-dashed) in the 2+3 topology.  b) The quality-of-fit variable
  (defined in equation~\ref{eq:qualityoffit}) for each endpoint along the flat direction. A representative pseudoexperiment is chosen for illustration. The solid black curve corresponds to the total quality-of-fit variable $Q$ for the pseudoexperiment.}
\label{fig:distributions2}
\end{figure}
\subsection{2+3 Topology}

We also study the effect of the flat direction for the 2+3 topology. In this
case, the kinematic distributions have only endpoints but no edges which are
poorly populated, especially at low statistics.

As for the 2+2+2 topology, we plot the true value of endpoints in figure \ref{fig:distributions2} for each
kinematic distribution and compare them with the predicted endpoints
for the hypothetical masses, varied along the flat direction. A representative pseudoexperiment is again chosen to illustrate the point. We see
that $(m^2_{23})_{max}$ and $(m^2_{31})_{max}$ are the only endpoints which are
sensitive along this direction, and consequently they drive the mass
measurement.

Since these two distributions have relatively shallow tails, the measured endpoint of a low statistics sample falls fairly short of the true endpoint, leading to a large bias in the mass measurement. From figure~\ref{fig:distributions2}, it is clear that the bias is towards smaller masses in this case.

\bibliography{Nbody}

\end{document}